\documentclass[twocolumn,10pt,superscriptaddress, reprint]{revtex4-1}

\usepackage{amsmath,amssymb,amsthm}
\usepackage{graphicx}
\graphicspath{{figures/}}
\usepackage{color}

\usepackage[utf8]{inputenc}
\newcommand{\jao}[1] {}
\renewcommand{\jao}[1] {{\color{red}{\textbf{JAO:} #1}}}

\usepackage{mathtools}
\usepackage{amsmath}
\usepackage{float}
\usepackage{xurl}

\renewcommand\[{\begin{equation}}

\renewcommand\]{\end{equation}}

\begin{document}
	\title{Size limits sensitivity in all kinetic schemes}
	\author{Jeremy A. Owen}
	\email{jaowen@mit.edu}
	\affiliation{Department of Physics, Massachusetts Institute of Technology, Cambridge, MA 02139, USA}
	\author{Jordan M. Horowitz}
	\email{jmhorow@umich.edu} 
	\affiliation{Department of Biophysics, University of Michigan, Ann Arbor, MI 48109, USA}
	\affiliation{Center for the Study of Complex Systems, University of Michigan, Ann Arbor, MI 48104, USA}
	\affiliation{Department of Physics, University of Michigan, Ann Arbor, MI 48109, USA}
	\date{\today}

\begin{abstract}
	Living things benefit from exquisite molecular sensitivity in many of their key processes, including DNA replication, transcription and translation, chemical sensing, and morphogenesis. At thermodynamic equilibrium, the basic biophysical mechanism for sensitivity is cooperative binding, for which it can be shown that the Hill coefficient, a sensitivity measure, cannot exceed the number of binding sites. Generalizing this fact, we find that for any kinetic scheme, at or away from thermodynamic equilibrium, a very simple structural quantity, the size of the support of a perturbation, always limits the effective Hill coefficient. This support bound sheds light on and unifies diverse sensitivity mechanisms, ranging from kinetic proofreading to a nonequilibrium Monod-Wyman-Changeux (MWC) model proposed for the \textit{E.~coli} flagellar motor switch, and represents a simple, precise bridge between experimental observations and the models we write down. In pursuit of mechanisms that saturate the support bound, we find a nonequilibrium binding mechanism, nested hysteresis, with sensitivity exponential in the number of binding sites, with implications for our understanding of models of gene regulation.
\end{abstract}

\maketitle

Sensitivity---the size of the response to a small perturbation---is a key figure of merit for performance on a number of tasks accomplished by living cells, including sensing chemical concentrations \cite{bialek_physical_2005, aquino_know_2016}, accurate signal transduction in cascades \cite{ferrell2014ultrasensitivity3}, molecular discrimination \cite{hopfield_kinetic_1974, murugan_speed_2012, murugan_discriminatory_2014}, and gene regulation \cite{estrada_information_2016, tran2018precision}. It is also a basic experimental observable, and so there is a long history of theoretical work connecting sensitivity measures to underlying mechanisms that could explain them---going back to Hill's realization that the sigmoidal binding curve of oxygen to hemoglobin \cite{Bohr} could be explained by binding to ``aggregations'' of hemoglobin \cite{Hill1910}. 

The set of known mechanisms that can underlie high sensitivity is very diverse, growing to include---in the last fifty years---nonequilibrium ones such as the ``futile cycle'' and kinetic proofreading \cite{hopfield_kinetic_1974, ninio_kinetic_1975}, whose study raises significant new challenges. Nevertheless, the success of a remarkably homogeneous modeling approach, rooted in chemical kinetics, makes possible a search for unifying principles---laws of sensitivity.

The prototypical sensitivity mechanism in biophysics is the cooperative binding of multiple copies of a ligand to a macromolecule. The probability or fraction of the fully bound state is frequently fit with a Hill function  \cite{ferrell2014ultrasensitivity1},
\[\label{hill}
f(x) = \frac{x^H}{K^H + x^H},
\]
where $x$ is the concentration of the ligand, $K$ is an effective dissociation constant, and the \textit{Hill coefficient}, $H$ quantifies the (logarithmic) sensitivity. Equation \eqref{hill} arises as an effective description in many different contexts, with $H$ depending in a complicated way on underlying details. However, in all cases of binding at thermodynamic equilibrium, there is a simple upper bound: the Hill coefficient cannot exceed the maximum number $n$ of ligands that can be bound at once. This limit on the sensitivity in terms of $n$ is purely \textit{structural}, being independent of all affinities and kinetic parameters.

The bound on the Hill coefficient is just one example of the many tight links between structure and function that hold at thermodynamic equilibrium. Out of equilibrium, these links are broken. The result is a huge increase in the parameter count of our models, because now not only are properties of system states (e.g.~free energies) important, but also the kinetics of all transitions between them. As we increasingly confront biological phenomena for which equilibrium models are inadequate, there is therefore a need for general results that constrain nonequilibrium function in terms of coarse properties, sidestepping this parametric complexity \cite{ahsendorf2014framework, zoller2020nonequilibrium, zoller2021eukaryotic}.

In this work, we show how the equilibrium bound on the Hill coefficient admits a vast generalization to nonequilibrium systems. We find that for any kinetic scheme, the logarithmic sensitivity of any steady-state observable to a perturbation---as quantified, for example, by a Hill coefficient---cannot exceed the size of the \textit{support} of the perturbation, a simple structural quantity we introduce: the support is the set of states that the system \textit{leaves faster} after the perturbation than before. The size of the support is always less than the number of system states---the size of the kinetic scheme.

The support bound on sensitivity applies to a large class of models---all continuous-time Markov chains, sometimes known as ``kinetic schemes'' or ``kinetic networks''---that are ubiquitous in biophysics, arising as the master equation of chemical reaction networks or as a coarse-grained description of the conformational dynamics of a single macromolecule. To illustrate the range of biological contexts in which the support bound applies, we show how it advances our understanding of a nonequilibrium Monod-Wyman-Changeux (MWC) -like model proposed for the \textit{E.~coli} flagellar motor \cite{tu_nonequilibrium_2008, wang_non-equilibrium_2017}, recovers known limits to molecular discrimination in kinetic proofreading \cite{murugan_speed_2012, murugan_discriminatory_2014}, and yields bounds on the accuracy of nonequilibrium chemical sensing \cite{harvey_universal_2020, hartich_nonequilibrium_2015}. In each of these examples, the support bound provides a way to go from experimental measurements of sensitivity to a concrete prediction about the underlying mechanism.

Finally, we apply the support bound to a class of models describing unordered, nonequilibrium, cooperative binding of a ligand (such as a transcription factor)---studied by prior authors \cite{estrada_information_2016, tran2018precision, park2019dissecting} in the context of the highly sensitive Hunchback-Bicoid system \cite{gregor2007probing} in \textit{Drosophila}. The support bound yields an upper bound on the Hill coefficient exponential in the number of binding sites, exceeding the limits identified by numerical search of the parameter space \cite{estrada_information_2016, tran2018precision}. We find that the exponential bound can in fact be achieved, by a nonequilibrium mechanism we identify and call nested hysteresis.

 \begin{figure*}
	\begin{center}
		\scalebox{0.55}{\includegraphics{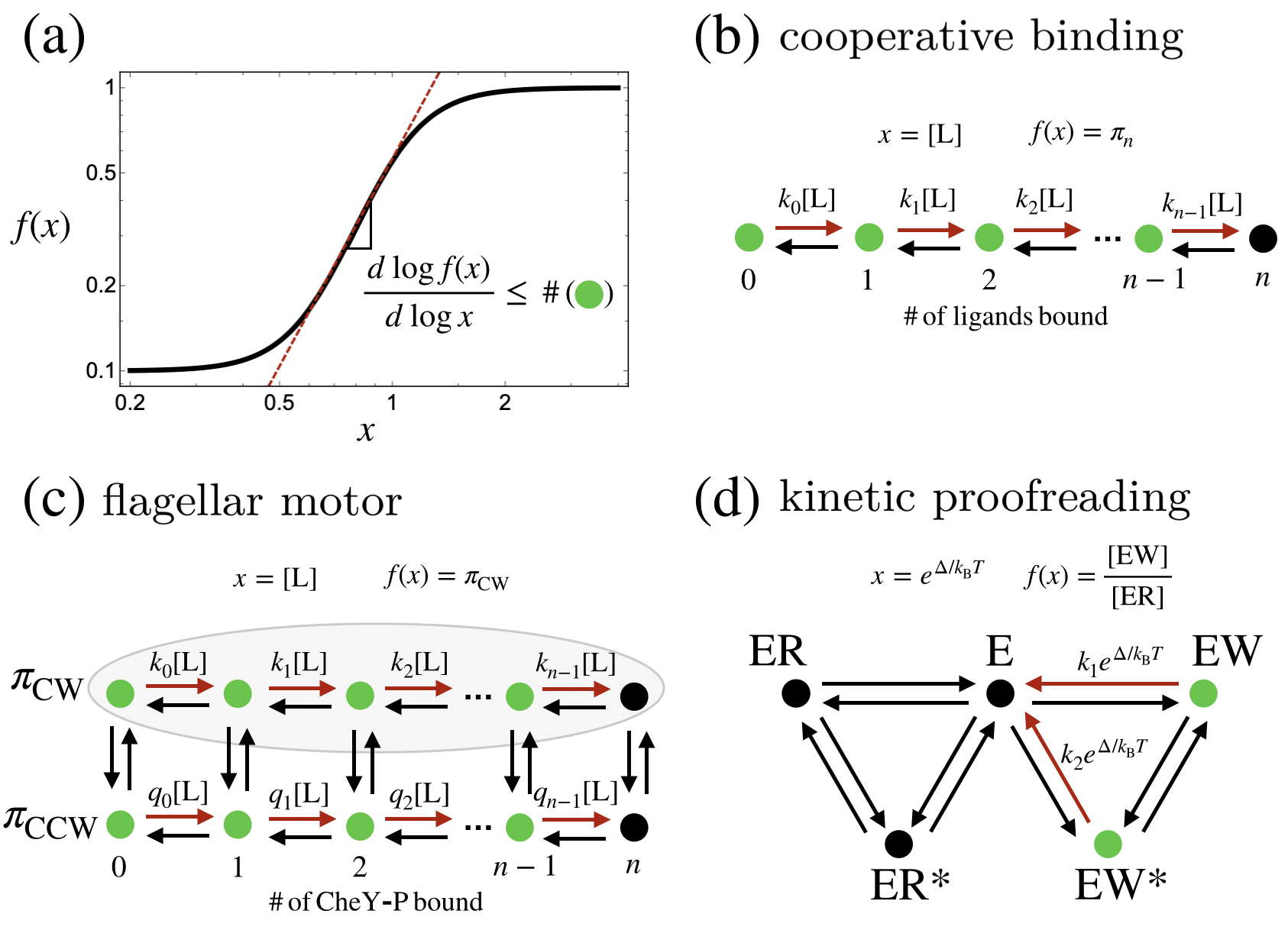}}
	\end{center}
	\caption{Illustration of our main result. (a) Example of a sensitive relationship between a parameter $x$ and a quantity $f(x)$. The slope on a log-log plot is a measure of sensitivity closely linked to the Hill coefficient. Our main result (inset equation) is that under general conditions this derivative is bounded by the \textit{size of the support of the perturbation} of $x$. In each example (b, c, or d), the graph $G$ of a kinetic scheme, to which our result applies, is shown. The transitions whose rates depend on $x$ are indicated in red. The support (green) consists of those states from which the red transitions \textit{leave}.} 
	\label{fig:main}
\end{figure*}

\section*{Definitions}   

\subsection*{Sensitivity measures}
The basic measure of sensitivity we focus on is the \textit{logarithmic sensitivity} of some quantity of interest $f(x)$ to a parameter $x$:
\[
\frac{d \log f(x)}{d \log x} = \frac{x}{f(x)} \frac{d f(x)}{ dx}.
\]
Here, we will briefly review the relationships between this derivative and other measures of sensitivity that might be reported or measured in an experiment, especially ``the Hill coefficient''.

If $f(x)$ were a Hill function (\eqref{hill}), then
\[
\label{hillderiv}
\frac{d \log f(x)}{d \log x} = H \left( \frac{K^H}{K^H + x^H} \right) = H(1-f(x)).
\]
We note two simple facts about this expression. First, the logarithmic sensitivity achieves its maximum value, $H$, when $x$ (and so $f(x)$) is very small. Second, at the midway point $x = K$, where $d f(x) / d \log x$ reaches its maximal value $H/4$, the logarithmic sensitivity is $H/2$. 

There is no guarantee that a function of interest will actually ``be'' a Hill function, but it is common nevertheless to report an ``effective Hill coefficient'', $H_{\text{eff}}$. There are in fact several \textit{distinct} quantities sometimes called the effective Hill coefficient. These different definitions all give $H_{\text{eff}} = H$ in the case of the Hill function, \textit{but in general they are not equivalent}. We will later see that the size of the support bounds them all.

First, suppose $f(x)$ is not a Hill function, but that it is known exactly, or at least, we can find its derivative. Then one approach is to define $H_\mathrm{eff}$ directly as the logarithmic sensitivity at some point, in analogy to how $H$ controls the sensitivity of the Hill function. For example,
\[
\label{hillcoeffderiv}
H_\mathrm{eff} = 2 \frac{d \log f(x) }{ d \log x }\Big\rvert_{x=x^*}
\]  
where $x^*$ is the value of $x$ at which $f(x)$ is halfway between the smallest and largest value it can assume. This definition has been used to quantify the sensitivity of non-Hill sigmoidal functions arising from theoretical models (e.g.~\cite{tu_nonequilibrium_2008, tran2018precision}).

In an experimental context, it is very common to fit a Hill function to data (e.g.~averaged observations) that are purported to reflect a functional relationship $f(x)$, and to report the fit parameter $\hat{H}$ as the Hill coefficient. Often this is informative, but as a matter of principle, two functions can have radically different derivatives even if the function values are very close everywhere (e.g.~if one function exhibits very high frequency but low amplitude oscillations). This means that, even if the fit is very good, relations based on analogy to \eqref{hillderiv}, such as that $\hat{H}/2 = d \log f(x) / d \log x$ at the midpoint, can fail dramatically. 

A different measure of sensitivity---the amplification of a fold-change in the input---provides the solution to this problem. Suppose that for some value $x_0$ of the input parameter $x$, scaling by a factor $a$ scales the output by $b$, so $f(ax_0) = b f(x_0)$. Then the quotient $\log(b)/\log(a)$ can be thought of as a discrete approximation of the derivative defining the logarithmic sensitivity. And if $f(x)$ is differentiable everywhere, then by the mean value theorem, there \textit{must} be a value $x^*$ of $x$ for which
\[
\label{fold-change}
\frac{d \log f(x)}{d \log x}\Big|_{x=x^*}  = \frac{\log b}{\log a}.
\]
This means that careful measurement of any two points on the input-output curve ($x$ versus $f(x)$) witnesses the (local, infinitesimal) logarithmic sensitivity \textit{somewhere}. Importantly---unlike in the case of fitting to a Hill function---if error in the measurements is very low, then they are also telling us the derivative for some value of $x$ very accurately.

Equation \eqref{fold-change} leads us to another common definition of the effective Hill coefficient \cite{goldbeter1981amplified}:
\[
\label{hillcoeff81}
H_{\text{eff}} = \frac{\log 81}{\log(S_{0.9}/S_{0.1})},
\]
where $S_{0.9}$ and $S_{0.1}$ are the values of the input variable (in our case, $x$) required to get $90\%$ and $10\%$ (respectively) of the maximum value of the output variable (in our case, $f(x)$). Note that \eqref{hillcoeff81} is like \eqref{fold-change} with $a = S_{0.9}/S_{0.1}$ and $b = 9$. It implies that \textit{somewhere} between $S_{0.1}$ and $S_{0.9}$ there is a logarithmic sensitivity of $H_{\text{eff}}/2$.

There is yet another common definition, specific to models of binding. Suppose $x$ is the concentration of a ligand and $\langle n_b \rangle(x)$ is the expected number of sites bound by a ligand out of a total of $n$ possible binding sites. It is common then, to take
\[
\label{hillcoeffbinding}
H_{\text{eff}} = \frac{d}{d \log x} \log\left(\frac{\langle n_b\rangle}{n-\langle n_b\rangle} \right),
\] 
or to report, as the Hill coefficient, the slope of a line fitted to $x$ versus $\left(\langle n_b\rangle\right)/\left(n-\langle n_b\rangle\right)$ data on a log-log plot.

As mentioned above, all these definitions of $H_{\text{eff}}$ are \textit{inequivalent} in general. For example, for totally noncooperative binding to $n = 2$ binding sites, we have $\langle n_b\rangle = 1 \times 2x/(1+x)^2 + 2 \times x^2/(1+x)^2 = 2(x/(1+x))$. In this case, \eqref{hillcoeffbinding} gives 1, as does \eqref{hillcoeff81}, if we take $f(x) = \langle n_b\rangle$. However, taking instead $f(x) = \left(x/(1+x)\right)^2$---the fraction of the time both sites are occupied---we get $H_{\text{eff}} \approx 1.17$ from \eqref{hillcoeffderiv} and $H_{\text{eff}} \approx 1.19$ from \eqref{hillcoeff81}.

\subsection*{Kinetic schemes}

Consider a system that may be modeled as undergoing stochastic transitions between a finite number of possible states $\{1,\dots, N\}$. Suppose the rate at which each transition occurs depends directly only on the current state, not on the past history of states or how long the system has resided in the current state. Then, the states of the system over time form a \emph{continuous-time Markov chain}. Models of this form are ubiquitous in nonequilibrium physics, chemistry, and biophysics, where they are known by many names including: kinetic schemes \cite{bialek2001stability}, kinetic networks \cite{vaikuntanathan2014dynamic}, Markov models, Markov jump processes, discrete-state kinetics \cite{iwahara2020discrete}, and linear framework graphs \cite{gunawardena2012linear, ahsendorf2014framework}. The main result we describe in this work applies to all such models.

In any continuous-time Markov chain, the probability $p_i(t)$ for a system to be found in state $i$ at time $t$, evolves according to the master equation:
\[
\label{master}
\frac{dp_i(t)}{dt} = \sum_{j=1}^N W_{ij} p_j(t),
\] 
where $W_{ij}$ is the rate of the transition from state $j$ to state $i$, and the diagonal entries $W_{jj} = -\sum_{i=1}^N W_{ij}$.

For a network of chemical reactions consisting only of \emph{monomolecular} reactions, \eqref{master} also arises as the rate equation describing the deterministic evolution of the vector of species concentrations \cite{mirzaev2013laplacian}. Both the stochastic and deterministic interpretations are common in applications, and we will elide this distinction by referring to any model whose dynamics are of the form \eqref{master} as a \emph{kinetic scheme}, though we will talk throughout as though we take the stochastic interpretation (e.g.~thinking of $p_i(t)$ as a probability).

To any transition rate matrix $W$ can be associated a weighted, directed graph $G$ whose vertices are the states of the system and whose directed edges represent allowed transitions, weighted by the transition rate. This graph $G$---which can be associated to \eqref{master} no matter its interpretation---plays a central role in the study of the scheme. In our figures, we will liberally use drawings of the graph $G$ to represent schemes. 

The \textit{cycles} of $G$ are important to stochastic thermodynamics, a theory which equips kinetic schemes with thermodynamic interpretations \cite{schnakenberg1976network, esposito2010three}. A cycle is a sequence of states and allowed transitions between them $1 \to 2 \to 3 \to \cdots i \to 1$, such that no two states in the sequence are the same except the first and the last, which coincide. Kinetic schemes describing systems at thermodynamic equilibrium must satisfy the principle of detailed balance \cite{lewis1925new, tolman1979principles}, which is equivalent to following condition on the rates around any cycle of $G$: 
\[
\label{db}
 \frac{W_{21}W_{32}\cdots W_{1i}}{W_{12}W_{23}\cdots W_{i1}} = 1.
\]
The support bound we establish \emph{does not} require detailed balance, and so it applies to models of nonequilibrium systems.

\subsection*{Observables and perturbations}

If the graph $G$ of a kinetic scheme is strongly connected (i.e.~there is a directed path of transitions from any state of our system to any other), then the solution $p_i(t)$ to \eqref{master} converges to a unique steady-state distribution $\pi$ satisfying
\[
\sum_i W_{ji} \pi_i = 0, \quad \quad \sum_i \pi_i = 1.
\]

For many systems whose dynamics are well-modeled by kinetic schemes, quantities typically measured experimentally are \text{averages} over observation times long enough that transients can be neglected, so that an average over the steady-state distribution is all that remains:
\[
\label{obs}
\langle A \rangle_\pi = \sum_i A_i \pi_i.
\]
Our focus in this work is how such ``steady-state observables'', or ratios of them, respond to changes in a parameter of interest $x$ that controls some of the transition rates $W_{ij}(x)$---i.e.~quantities of interest will always be of the form $f(x) = \langle A \rangle_\pi$ or $f(x) = \langle A \rangle_\pi / \langle B \rangle_\pi$.  Note that, since the steady-state distribution $\pi$ is uniquely determined given the matrix of transition rates $W$, we can view it as a function of $W$, and so by extension as a function of the parameter $x$, $\pi(W(x))$. This means that steady-state observables are also functions of $x$.

We will consider only \emph{positive} observables, which are sums of the form \eqref{obs} for which each of the $A_i$ is nonnegative and at least one is positive (if this is not true, we can add a constant to all the $A_i$ to make it so). We also suppose throughout that the parameter $x$ is positive. We further restrict attention to the case where $x$ simply \emph{multiplicatively scales} some of the transition rates. In other words, there are two kinds of rates: those that do not depend on $x$, and those that do, whose $x$-dependence is always of the form $W_{ij} = k_{ij} x$, where $k_{ij}$ is a positive number independent of $x$. In most of our examples, $x$ will be the concentration of a chemical species $L$ and some of the transitions in a kinetic scheme will correspond to the binding of $L$, so that the law of mass action leads to a linear dependence on $x$. 

The \textit{support of the perturbation} of $x$ is defined as the set of states (vertices in $G$) that have at least one outgoing transition (directed edge) that depends on $x$. The support consists of exactly those states whose \textit{exit rates} depend on (and are increasing in) $x$. Note that the size of the support cannot exceed the total number of states $N$ in the scheme.

\section*{Results}

\subsection*{The support bound}

We are now ready to state our main result. If $A$ and $B$ are positive observables of a kinetic scheme, and $m$ is the size of the support of the perturbation of $x$, then:

\begin{equation}\label{supportbound}
\left|\frac{d \log \langle A \rangle_\pi / \langle B \rangle_\pi}{d \log x}\right| \leq m.
\end{equation}

We call this inequality the support bound. Our proof of it, which we give in the Materials and Methods, is an application of the Markov chain tree theorem \cite{tutte1948, hill1966, shubert1975, schnakenberg1976network, leighton1986, mirzaev2013laplacian}, which gives, for any kinetic scheme, an explicit algebraic expression for the steady state  $\pi$ in terms of all the transition rates. This result \eqref{supportbound} is related to prior results in the Markov chain literature, especially those by O'Cinneide \cite{o1993entrywise}  and Takahashi \cite{takahashi1973effects}, which effectively give bounds on sensitivities in terms of total number of states $N$. It is also closely related in spirit to the results of Wong et al.~\cite{wong_structural_2018}, who apply the Markov chain tree theorem to  find structural conditions for the emergence of the Michaelis-Menten formula from general kinetic schemes. 

The inequality \eqref{supportbound} also serves as a companion result to those of our own prior work \cite{owen2020universal}, which aimed to understand nonequilibrium response by decomposing perturbations into sums of ``vertex'' and ``edge'' perturbations. In that language, the support bound provides a shortcut to the ultimate limits of sensitivity to vertex and symmetric edge perturbations, when thermodynamic constraints are completely loosened. And for combinations of asymmetric edge perturbations, the support bound usually provides a tighter bound than our prior results.

A useful corollary of \eqref{supportbound} follows from taking the observable $A$ to be the indicator function of a subset $X$ of states and $B$ to be the indicator function of the complement of $X$. In this case, $\langle A\rangle = \pi_X$ is the steady-state probability of finding the system in one of the states of $X$ and $\langle B\rangle = 1-\pi_X$, leading to the result
\[\label{boundforp}
\left|\frac{d \log \pi_X }{d \log x}\right| \leq m (1-\pi_X).
\]

We want to emphasize the generality of support bound \eqref{supportbound} and its corollary \eqref{boundforp}---they hold at steady state for any kinetic scheme, whether it respects or breaks detailed balance, for any positive observables $A$ and $B$, for any subset of states of interest $X$, and for any parameter $x$ that scales some transition rates. The right hand side of the support bound is independent of all aspects of the underlying kinetic scheme, except for the size $m$ of the support of the perturbation---which is the number of system states that have at least one exit transition whose rate is scaled by $x$.

Note the similarity of the right hand side of \eqref{boundforp} to the derivative of a Hill function, \eqref{hillderiv}. A key consequence of our results is that \textit{the effective Hill coefficient---in all the different incarnations discussed earlier---is always bounded by $m$}. To see this for the ``binding'' definition \eqref{hillcoeffbinding} of $H_\mathrm{eff}$, we can apply \eqref{supportbound}. For the definition \eqref{hillcoeffderiv}, \eqref{boundforp} is sufficient. It is hardest to see that the ``non-local'' definition \eqref{hillcoeff81} is bounded by $m$. We show this in Appendix~\ref{global_hill_bounded}.   

\subsection*{Ordered binding schemes}

To illustrate our result, we begin with a familiar example where we can compare the support bound to a transparent, exact formula for sensitivity. Consider a ligand $L$ with concentration $x$, and suppose that up to $n$ copies of the ligand can bind \textit{in order} to a receptor. Equivalently, we can imagine that the ligand can bind in any order but that rates of binding and unbinding depend only on the number bound (e.g.~binding sites are indistinguishable). These scenarios both give rise to the kinetic scheme illustrated in Figure \ref{fig:main}(b). By the law of mass action, transition rates for binding events are proportional to the ligand concentration $x$.

There are $N = n+1$ states in total, but the support of the perturbation is of size $m = n$. So the support bound implies that, for example, if $\pi_n$ is the steady-state probability that the maximum number $n$ of ligands is bound, then 
\[
\frac{d \log \pi_n}{d \log x} \leq n (1 - \pi_n)
\]
In this case, we also have the exact formula for sensitivity:
\[
\frac{d \log \pi_n}{d \log x} = n - \langle n_b \rangle_\pi,
\]
and so the support bound is saturated when $\langle n_b \rangle_\pi = n \pi_n$, in other words, in the limit of extreme cooperativity---either no ligands are bound or they all are.

\subsection*{Any detailed balanced scheme} 
Remarkably, for any binding scheme the satisfies detailed balance, the sensitivity to the ligand concentration $x$ is given by a simple expression. For any detailed balanced scheme (including \textit{all} models describing thermodynamic equilibrium), if $X$ is a set of states of interest, and $\overline{X}$ is the set of states not in $X$, then we have (Appendix~\ref{proof_eqbound})
\[
\label{eqbound}
\frac{d \log \pi_X}{d \log x} =\underbracket[0.187ex]{\left[ \langle n_b \rangle_X - \langle n_b \rangle_{\overline{X}} \right]}_{H_\text{eff}} (1-\pi_X), 
\]
where $\langle n_b \rangle_X$ (resp.~$\langle n_b \rangle_{\overline{X}}$) is the expected number of ligands bound, conditional on the system being found in one of the states of $X$ (resp. $\overline{X}$).

The right hand side of \eqref{eqbound} cannot exceed $n(1-\pi_X)$, where $n$ is the maximum possible number of ligands that can be bound. Therefore, comparing to \eqref{hillderiv}, we see how this formula \eqref{eqbound} refines the observation that, at thermodynamic equilibrium, the effective Hill coefficient cannot exceed the number of binding sites. This result will prove important in the next section, where we will compare it to the support bound, which holds both in and out of equilibrium.

Equation \eqref{eqbound} is closely related to the well-known theorems relating response to fluctuations at thermodynamic equilibrium (see e.g.~Section 7.1 in \cite{zwanzig2001nonequilibrium}) and similar expressions have also appeared in a biophysical context (Eq.~8.2 in \cite{wyman1964linked}). A physical argument for \eqref{eqbound} can be made using the grand canonical ensemble, which tells us that probability of any system state in which $n_b$ ligands are bound is proportional to $\exp(\beta \mu n_b) \sim \exp(n_b \log x)$, where $\beta\mu$ is the chemical potential of the ligand, and the proportionality ``$\sim$'' elides factors that do not depend on the ligand concentration, $x$. Differentiating the resulting expressions yields the result. A complete proof of  \eqref{eqbound}, directly from the principle of detailed balance, can be found in Appendix \ref{proof_eqbound}.

\begin{figure*}
	\begin{center}
		\scalebox{0.45}{\includegraphics{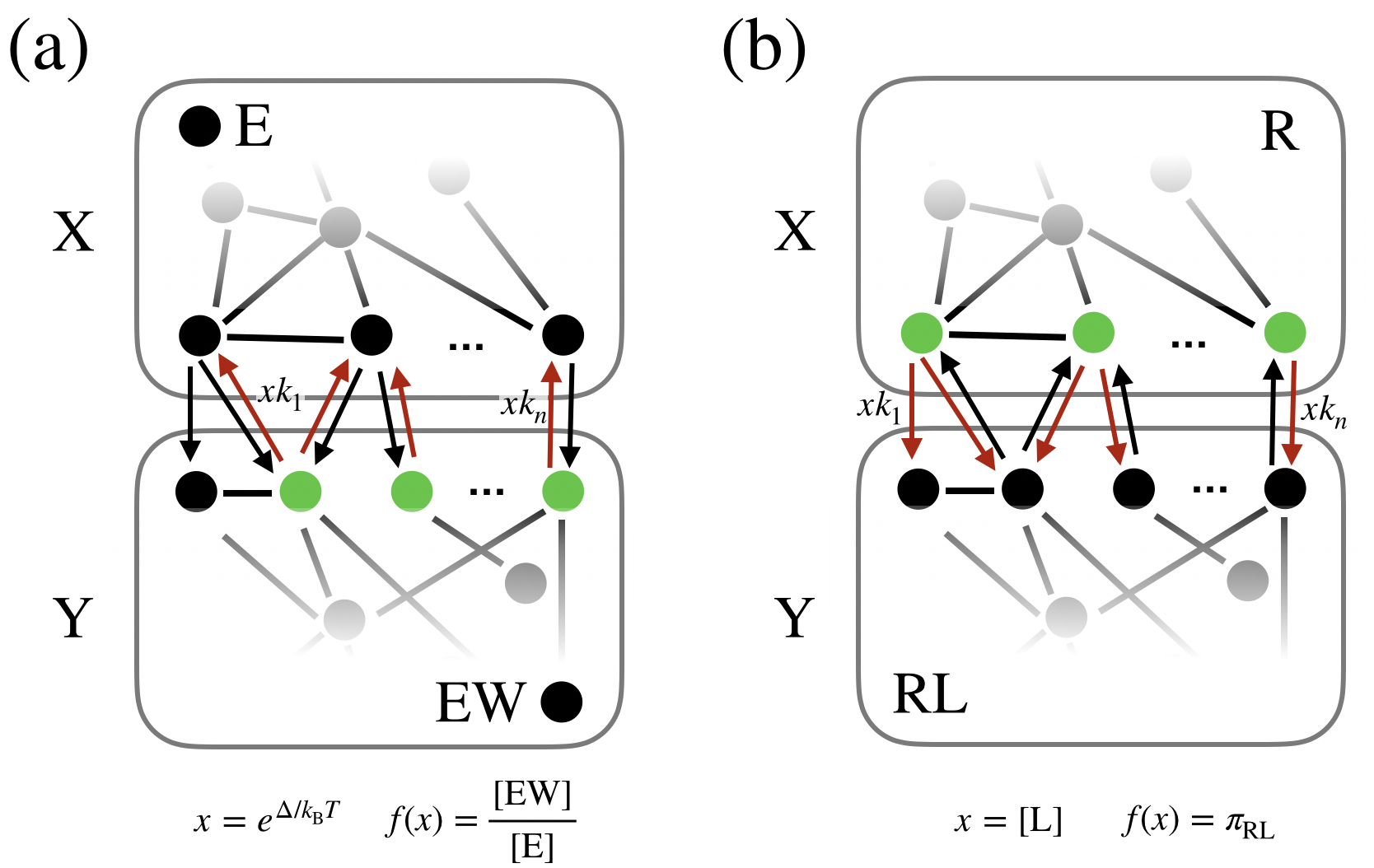}}
	\end{center}
	\caption{Common structure of sensing and proofreading models. (a) Murugan's generalized proofreading scheme \cite{murugan_discriminatory_2014}, where the key assumption is that there is a ``discriminatory fence'' dividing the states into two halves, and every transition depending on the energy difference $\Delta$ crosses this fence. (b) The general receptor model used in \cite{harvey_universal_2020} to study the sensing of a ligand concentration has the same structure, with the separation between the bound and unbound states of the receptor playing the role of the ``fence''.}
	\label{fig:fence}
\end{figure*}

\subsection*{Nonequilibrium MWC and the flagellar motor}

Now we turn to bacterial chemotaxis, where the support bound sheds light on the (possibly nonequilibrium) mechanism underlying the sensitive directional switching of the flagellar motor.

In the chemotaxis system of \textit{E.~coli}, an array of receptors senses the chemical environment of the cell, and controls the intracellular concentration of the phosphorylated protein $\text{CheY-P}$. In turn, the $\text{CheY-P}$ concentration controls the direction of rotation of the flagellar motors of \textit{E.~coli}---determining whether the bacterium ``runs'' or ``tumbles''. The relationship between $[\text{CheY-P}]$ and the fraction of the time a motor rotates clockwise, $\pi_\text{CW}$, is known to be an extremely sensitive one, with studies \cite{kuo1989multiple, scharf1998control, alon1998response, cluzel_ultrasensitive_2000} over time reporting increasingly large Hill coefficients, as experimental techniques have more fully isolated a single motor's ``input-output relation''. Recent measurements, due to Yuan and Berg \cite{yuan_ultrasensitivity_2013}, found $H \approx 21$.  

The underlying mechanism generating this sensitivity is unknown, but is thought to involve the binding of $\text{CheY-P}$ to some of the $\sim34$ FliM protein subunits of the motor, promoting clockwise rotation. There have been several equilibrium models of cooperative binding proposed for this, including, notably, the Ising-like conformational spread model \cite{duke_conformational_2001}.

But for any equilibrium model, including the conformational spread model, \eqref{eqbound} predicts that the Hill coefficient for directional switching is given by the difference in the mean number of bound $\text{CheY-P}$ molecules in the clockwise (CW) and counterclockwise (CCW) rotation states. Fukuoka et al. \cite{fukuoka_direct_2014} measured a quantity very much like this---finding an average of 13 $\text{CheY-P}$ molecules are bound when the motor rotates CW, compared to an average of only 2 bound during CCW rotation. This measurement may be mixed up with intrinsic fluctuations of $[\text{CheY-P}]$, but even allowing for this, the difference of these numbers $\sim 11$ should still \textit{exceed} the Hill coefficient (Appendix \ref{eqbound_fukuoka}), contradicting the finding $H \approx 21$ of Yuan and Berg. A nonequilibrium mechanism is needed to reconcile these observations.

Other lines of evidence, including observations of the statistics of the time spent in the CW or CCW states between switching events, also point to a nonequilibrium mechanism \cite{wang_non-equilibrium_2017, tu2017driven}. Tu \cite{tu_nonequilibrium_2008}, proposed a simple nonequilibrium model for directional switching. The model, illustrated in Figure \ref{fig:main}(c), is a kinetic scheme with the structure of an MWC model---coupling the binding of $n$ ligands (CheY-P) to a global (i.e.~concerted) transition between the two motor states (CW or CCW)---except that detailed balance is broken. 

Tu assumed a particular form for the rate constants in the model, but here we relax the choice of rate constants, and ask what sensitivity is possible in models with this general, ``nonequilibrium MWC'' structure. By counting the green states in Figure \ref{fig:main}(c), we see that $m = 2n$ for models in this class. Therefore, the support bound constrains the sensitivity of the clockwise bias $\pi_\text{CW}$ to changes in the CheY-P concentration, $x = [\text{CheY-P}]$, as
\[
\label{chemotactic_bound}
\frac{d \log \pi_\text{CW}}{d \log [\text{CheY-P}]} \leq 2 n (1-\pi_\text{CW}),
\]
or $H_\text{eff} \leq 2n$. This bound can be approached arbitrarily closely in an appropriate limit of transition rates (Appendix \ref{chemotaxis_sat}). In fact, $\pi_\mathrm{CW}$ can be seen to approach a Hill function with $H = 2n$, $\pi_\mathrm{CW}(x) \to x^{2n}/(K^{2n}+x^{2n})$, saturating \eqref{chemotactic_bound}. 

The sensitivity optimizer we find is qualitatively similar to Tu's choice of parameters---positing a nonequilibrium enhancement of the $\mathrm{CW} \to \mathrm{CCW}$ transition when few CheY-P are bound and an enhancement of $\mathrm{CCW} \to \mathrm{CW}$ transition when many CheY-P are bound. Such an enhancement might arise from coupling of the switching transitions to the torques generated by the motor \cite{wang_non-equilibrium_2017, wang2021dynamics}. 

In models of this form, $n$ is the difference between the largest and smallest possible number of bound ligands. For a model of the flagellar motor, the simplest interpretation is that $n \approx 34$, the number of FliM subunits. In this case, our result provides only a very loose bound $H_\text{eff} \leq 2\times 34 = 68$. However, Fukuoka et al.~found that very high FliM occupancies were rare. If it were the case that the number of bound CheY-P molecules were constrained to \textit{never} leave the range 2 to 13, then we could take $n = 13 - 2 = 11$. $2n = 22$ would then be suggestively close to the Hill coefficient measured by Yuan and Berg \cite{yuan_ultrasensitivity_2013}. Unfortunately, it is possible that transient passage through rare states could have an outsized effect on sensitivity. All \eqref{chemotactic_bound} says is that to explain a Hill coefficient of 21 using a model of this form, it is necessary to allow for a range of least $n = 11$ in the number of ligands bound.

Importantly, $n$ is \emph{not} the difference in the means $\langle n_b \rangle_\text{CW} - \langle n_b \rangle_\text{CCW}$, and we are not saying that $H_\text{eff} \leq 2 \left(\langle n_b \rangle_\text{CW} - \langle n_b \rangle_\text{CCW}\right)$ always. This is not true---in fact, it is possible in a nonequilibrium MWC model to see a Hill coefficient of $n$, $\pi_\mathrm{CW}(x) \approx x^{n}/(K^{n}+x^{n})$, even if $\langle n_b \rangle_\text{CW} - \langle n_b \rangle_\text{CCW} = 0$ (Appendix \ref{chemotaxis_sat}). Additionally, we want to emphasize that the concordance between sensitivity measurements and the number of ligands bound is not the only desideratum for a model of flagellar motor switching, and we are not putting forward the optimizer of \eqref{chemotactic_bound} that we describe in Appendix \ref{chemotaxis_sat} as \emph{the} motor mechanism.

\subsection*{Proofreading and sensing}

Next, we turn to the example of kinetic proofreading (KP) \cite{hopfield_kinetic_1974, ninio_kinetic_1975}. We will see how the support bound limits the accuracy of molecular discrimination in terms of the number of proofreading ``steps'', recovering and very slightly refining the limits found by Murugan et al.~\cite{murugan_discriminatory_2014} for general KP schemes. Then, we turn to the problem of sensing chemical concentrations, where models with exactly the same structure arise \cite{hartich_nonequilibrium_2015}. In this case, the support bound yields new constraints on the accuracy of chemical sensing by a nonequilibrium receptor \cite{harvey_universal_2020}, in terms of the number of distinct receptor states that can bind ligand.

To frame the discussion of KP, consider an enzyme $E$ that can bind either of two very similar substrates present at equal concentrations---a ``right'' one $R$ and a ``wrong'' one $W$. Suppose the enzyme-substrate complex $EW$ has a free energy larger than that of $ER$ by a small amount $\Delta k_\mathrm{B} T$. Then, at thermodynamic equilibrium, the so-called error fraction $\eta \equiv [EW]/[ER]$ equals the Boltzmann factor $\eta = b \equiv \exp(-\Delta)$. Typically, it is supposed that irreversible product formation can occur from the intermediates $ER$ or $EW$, and it is desired that this mostly occur from $ER$ rather than from $EW$. If product formation is slow compared to other rates, then this is equivalent to desiring that $\eta$ be small.

Out of equilibrium, one might imagine it is easy to lower the error fraction by selectively coupling the binding of $W$ to nonequilibrium processes. But this is unsatisfactory, because it simply ``passes the buck'' of discrimination to the processes that do that coupling. KP is a kinetic scheme, illustrated in Figure \ref{fig:main}(d), that can instead \textit{amplify} the effect of the small energy difference $\Delta$, while otherwise treating $R$ and $W$ the same. The degree of this amplification can be quantified by a \textit{sensitivity}---the ``discriminatory index'' $\nu$, introduced by Murugan \cite{murugan_discriminatory_2014}:
\[
\nu = -\frac{d \log([EW]/[ER])}{d \Delta} = -\frac{d \log([EW]/[E])}{d \Delta},
\]
which we note is of a form we can constrain with the support bound \eqref{supportbound}, taking $x = \exp(\Delta)$. For the scheme in Figure \ref{fig:main}(d), the size of the support is the number of bound states of $W$ that can dissociate, so $m = 2$, recovering the well-known fact that $\nu \leq 2$. This upper limit  $\nu \to 2$ can indeed be approached for an appropriate choice of transition rates, corresponding to Hopfield’s result that the error fraction can approach $b^2$.

The support bound also recovers the limits of discrimination in more general KP schemes \cite{murugan_speed_2012, murugan_discriminatory_2014}. When computing $\nu$ for complicated schemes, it is helpful to restrict attention to the part of the graph on which $\nu$ depends, which is just the reactions involving bound states containing $W$ (neglecting the totally analogous ones involving $R$). Showing just half of a symmetric kinetic scheme, the class of generalizations we consider are illustrated in Figure \ref{fig:fence}(a). We consider \textit{any} scheme whose states can be divided into two sets $X$ and $Y$, such that the transitions depending on the parameter $x$ are exactly those crossing $Y$ to $X$.

The KP interpretation of Figure \ref{fig:fence}(a) is that somewhere in $X$ there is a state that represents the unbound enzyme $E$, and somewhere in $Y$ there is a state that represents a bound state $EW$. Somewhere along the paths between these states, there are transitions that cross, in Murugan's language, a ``discriminatory fence'' \cite{murugan_discriminatory_2014} and depend on  $x = \exp(\Delta)$. Murugan focused on the number $c$ of transitions crossing this ``fence'', and found $\nu \leq c$. The support bound \eqref{supportbound} tells us instead to count the number $m$ of ``boundary states'' from which these crossing transitions emanate from, yielding $\nu \leq m \leq c$. Saturation of this bound is possible by the ``ladder''-like multi-step proofreading schemes discussed in \cite{murugan_speed_2012}.

Now we turn to a seemingly different example, which is the challenge faced by cells of sensing chemical concentrations. Harvey et al.~\cite{harvey_universal_2020} sought to understand universal constraints on sensing by studying a model of a general nonequilibrium receptor, illustrated in Figure \ref{fig:fence}(b), with exactly the same structure as Murugan's general KP scheme. In context of their work, $X$ corresponds to states in which the receptor is not bound to a ligand (``nonsignaling states''), and  $Y$ corresponds to those when it is (``signaling state''), and $x = c$, the concentration of the ligand being sensed. Again the support bound will be of use, although to bound the sensitivity of a different observable (not the discriminatory index).

The key question in sensing is how well $c$ can be estimated. The accuracy is related to a sensitivity that we can constrain using the support bound. To see how this works, suppose a cell's surface is covered by a number $R_T$ of identical, non-interacting receptors modeled by a scheme of the form in Figure \ref{fig:fence}(b), and they relax to a steady-state distribution over their states subject to fixed external ligand concentration $x = c$. Now consider the number $r$ of the receptors that are in a ``signaling state'' (one of the states in the set $Y$), at one particular instant.  For the sake of example, this will be our ``readout'' \cite{govern2014energy} from which we wish to construct an estimate $\hat{c}$ of $c$. The mean of $r$ is a function of $c$ given by $f(c) = R_T \pi_Y(c)$, and if it is invertible, we could take $\hat{c} = f^{-1}(r)$. The error in an estimate so constructed can, under certain assumptions, be approximated as
\[
\epsilon_{\hat{c}}^2 \equiv \frac{\operatorname{Var} \hat{c}}{c^2} \approx \frac{\operatorname{Var} r}{R_T^2\left(c \frac{d \pi_Y}{d c}\right)^2}.
\]

Now, $r$ is a binomial random variable, so $\operatorname{Var} r = R_T \pi_Y(1-\pi_Y)$, and the derivative in the denominator we can bound in terms of the size of the support, $m$,
\[
c \frac{d \pi_Y}{d c} = \frac{d \pi_Y}{d \log c} \leq m  \pi_Y(1-\pi_Y)
\]
leading to a lower bound on the sensing error in terms of the support:
\[
\label{sensing_bound}
\epsilon_{\hat{c}}^2 \geq \frac{1}{R_T m^2 \pi_X (1-\pi_X)} \geq \frac{4}{R_Tm^2}.
\]
The larger the support, the higher is the achievable sensitivity and the lower is the achievable sensing error.

\begin{figure*}
	\begin{center}
		\scalebox{0.55}{\includegraphics{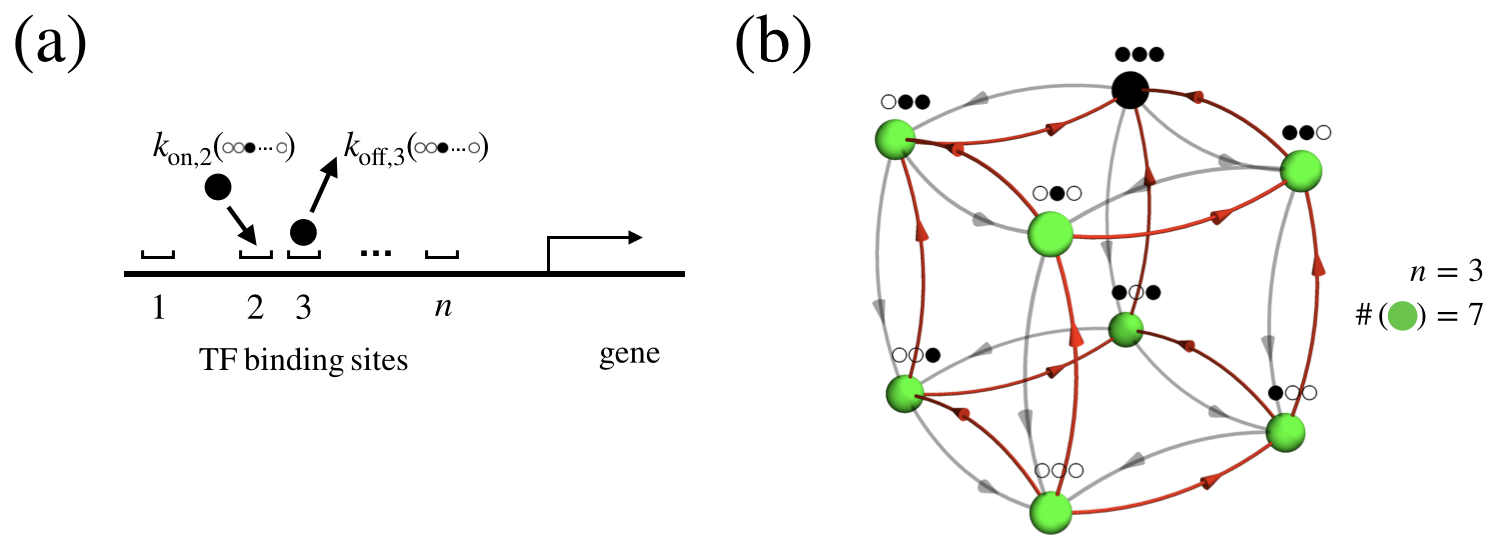}}
	\end{center}
	\caption{Unordered binding. (a) A gene may be regulated by the binding of transcription factors (TFs) to some number $n$ of sites. The most general case is that of unordered binding to distinguishable sites. The TFs might bind in any order, and the rates $k_\mathrm{on}$ and $k_\mathrm{off}$ of binding and unbinding may be different for each site and depend on the occupancy state of all the other sites. (b) The graph of states and transitions for unordered binding of $n=3$ copies of a ligand.}
	\label{fig:unordered}
\end{figure*}

\subsection*{Unordered binding and nested hysteresis}

Finally, we turn to the application of the support bound to models in which identical ligands bind, in any order, to $n$ \emph{distinguishable} binding sites, and in which detailed balance may be broken in the binding and unbinding rates. In this case, we will see that the support bound gives a limit on sensitivity that is exponential in $n$. We find that this remarkable sensitivity can in fact be achieved, by a simple mechanism we call \textit{nested hysteresis}.

The motivating example in this section will be the regulation of a gene by the binding of copies of a transcription factor (TF) to multiple sites along a DNA molecule (Figure \ref{fig:unordered}(a)). Gene expression can be strikingly sensitive to TF concentration. For example, in the \textit{Drosophila} embryo, an exponentially decaying spatial gradient of the TF called Bicoid is transformed into a sharply sigmoidal pattern of Hunchback gene expression across the embryo \cite{driever1988bicoid, struhl1989gradient, houchmandzadeh2002establishment}. These observed patterns can be fit to a Hill function with $H \sim 5 - 7$ \cite{gregor2007probing, estrada_information_2016, tran2018precision, park2019dissecting}. Many authors have proposed to explain this as a consequence of equilibrium cooperative binding to $5 - 7$ Bicoid binding sites \cite{driever1989determination, ma1996drosophila, gregor2007probing, xu2015combining}, but this picture is at least clouded by recent theory and experiments which found effects of binding site deletions that were contrary to equilibrium expectations \cite{estrada_information_2016, park2019dissecting}. And indeed, especially in eukaryotes, there are many avenues by which energy may be expended in gene regulation, breaking detailed balance \cite{wong2020gene}. 

Inspired by this example, Estrada et al.~\cite{estrada_information_2016} asked what relationships between TF concentration and gene expression could arise from the binding of TFs to distinguishable sites, in any order, without assuming detailed balance  (Figure \ref{fig:unordered}(a)). In unordered binding, each of $n$ binding sites can be occupied or not, independently of the others, so there are $2^n$ possible states in the kinetic scheme. The allowed transitions are those involving the binding or unbinding of single TF molecule, resulting in a hypercube graph of states and transitions (illustrated in Figure \ref{fig:unordered}(b) for $n = 3$). As usual, the binding transitions are assumed to have rates linear in the TF concentration, $x$, and all other transition rate are independent of $x$.

Finally, Estrada et al.~\cite{estrada_information_2016} supposed that the level of gene expression was, in the language of our work, a positive steady-state observable (e.g.~proportional to the steady-state probability of all sites being bound, or to the average number of ligands bound). They then searched the parameter space of their models numerically---seeking the extremes of shape of the resulting functional relationships between $x$ and gene expression---and compared these shapes to those of Hill functions. The subsequent work of Tran et al.~\cite{tran2018precision} performed similar explorations.

What does the support bound say about sensitivity in such models? The support of the perturbation of $x$ in the case of unordered binding consists of every single state of the system, except for the fully bound state, in which every binding site is occupied and no more binding can occur. Therefore, the size of the support is $m = 2^n - 1$ and the support bound yields, for example,
\[
\label{hypercubebound}
\frac{d \log \pi_\mathrm{all}}{d \log x} \leq \left(2^n - 1\right)(1 - \pi_\mathrm{all}),
\]
where $\pi_\mathrm{all}$ is the steady-state probability of the fully bound state. Or as discussed earlier, we may also say, more roughly, $H_\mathrm{eff} \leq 2^n - 1$. Can such exponential-in-$n$ sensitivity actually be achieved? We find that it can be, by a simple mechanism---nested hysteresis---that we describe in the next paragraph. And in fact, by a small further elaboration of the mechanism, it appears that $\pi_\mathrm{all}$ (viewed as a function of the ligand concentration $x$), can be made as close as desired to a Hill function with $H = 2^n - 1$.

There are two key ingredients in nested hysteresis. Suppose the binding sites are numbered: $1, \dots, n$. The first ingredient is a hierarchy of timescales, such that binding and unbinding to each successive (higher-numbered) site is much slower than to the (lower-numbered) one before. The exponential-in-$n$ sensitivity will depend on very strong separation of timescales in this hierarchy. The second ingredient is a simple rule restricting when binding and unbinding can occur---\textit{binding} to a site can only happen when all the lower-numbered ones are \textit{bound}, and \textit{unbinding} from a site can only happen when all lower-numbered ones are \textit{unbound}. When binding or unbinding at a site can occur, they occur at some rate and we suppose the ratio of these rates equals $x$ (as if working in units where the dissociation constant equals one). 

\begin{figure*}
	\begin{center}
		\scalebox{0.6}{\includegraphics{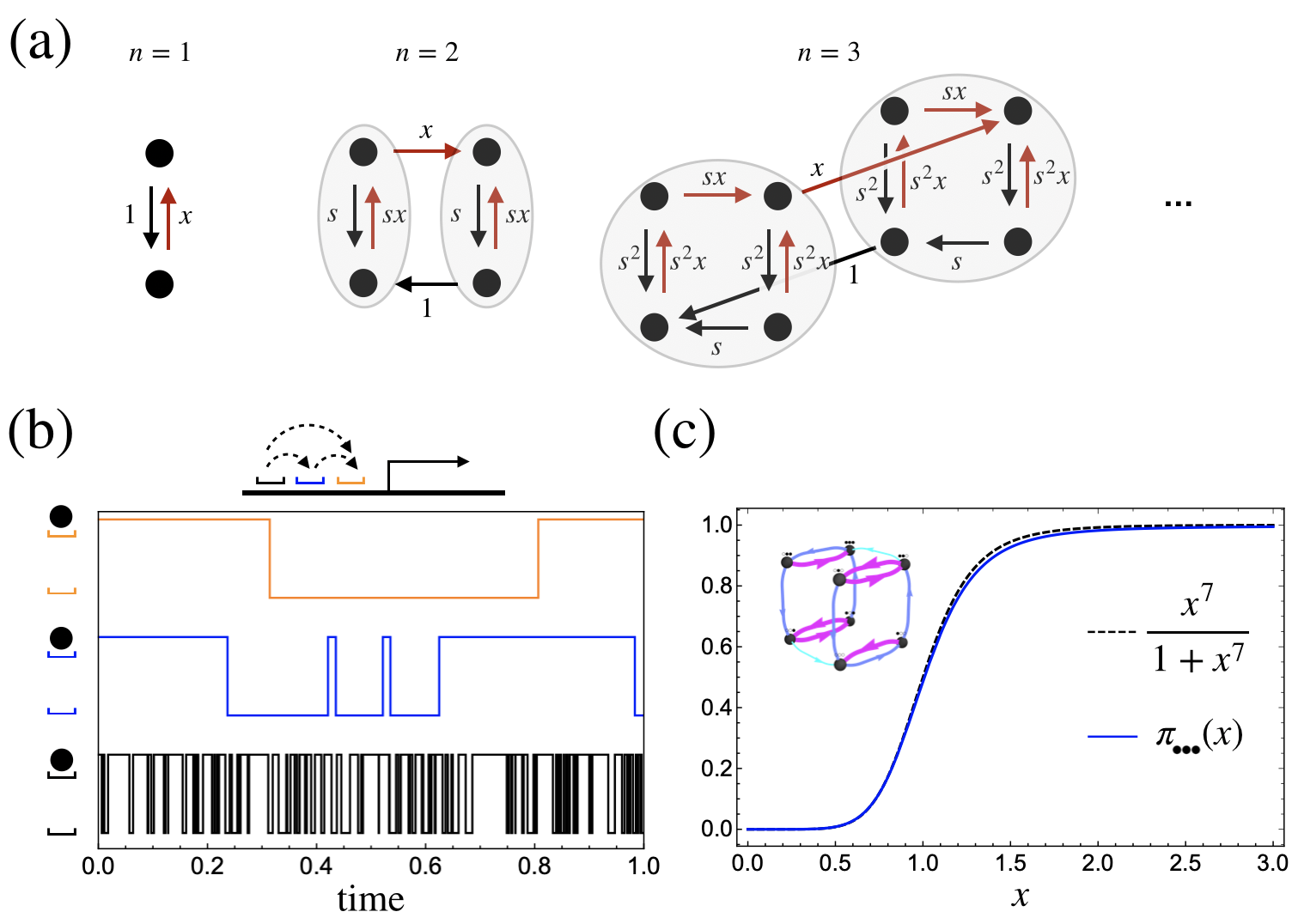}}
	\end{center}
	\caption{Nested hysteresis. (a) Iterative construction of the kinetic scheme of nested hysteresis, generalizable to any value of $n$. In each diagram, the gray ovals indicate the subsystems (corresponding to binding and unbinding to the first $n-1$ sites), which are assumed to relax much faster than the other transitions. (b) Occupation over time for each of $n = 3$ binding sites (black, blue, and orange), in a particular stochastic realization of nested hysteresis (parameters: $n=3$, $s=10$, $x=2$). Dotted arrows in the schematic indicate the pattern of influences between the binding sites. (c) Illustration of a parameter choice (inset, color indicates rate---magenta $ =10^4$, blue $=100$, cyan $=1$) for which the dependence on $x$ of the probability of the fully bound state approaches a Hill function with $H=7$ very closely.}
	\label{fig:nh}
\end{figure*}

These rules gives rise to a nested structure, where the dynamics at each binding site depend \emph{only} on lower-numbered ones. The iterative construction of a kinetic scheme realizing this mechanism for any $n$---incorporating an explicit scale factor $s$ giving rise to the required timescale separation in the limit $s \to \infty$---is illustrated in Figure \ref{fig:nh}(a). Typical stochastic dynamics of this scheme for $n = 3$ and a finite value of $s$ are shown in Figure \ref{fig:nh}(b), illustrating its hallmarks---the hierarchy of timescales, and the dependence of the dynamics of higher-numbered sites on the occupancy of the lower-numbered ones.

In the limit of strong timescale separation, we can analytically find the steady-state distribution of nested hysteresis. We give here an intuitive argument, and provide more careful arguments in Appendix \ref{nested_hysteresis_details}. To begin, we start by considering the first binding site. This site is independent of all the others, and the ratio of the binding rate to the unbinding rate is $x$, so the steady-state probability that the first site is bound is
\[
\pi\left(\text{site 1 is bound}\right) = \frac{x}{1+x}.
\]
By assumption, binding to the second site can only happen when the first is bound. And, since (again, by assumption) there is a strong timescale separation between these two sites, this amounts to an effective rate of binding to the second site of $k x \times \pi\left(\text{site 1 is bound}\right) = \frac{kx^2}{1+x}$, where $k$ is a constant that will drop out. Unbinding happens at an effective rate of $k \times \pi\left(\text{site 1 is not bound}\right) = \frac{k}{1+x}$. From this it follows that
\[
\pi\left(\text{site 2 is bound}\right) = \frac{x^2}{1+x^2}.
\]

Importantly, it is also a consequence of the timescale separation that the sites behave as though they are independent at steady state, in the sense that the probability both are bound is the product of the probabilities that each one is, so that
\[
\pi\left(\text{sites 1 and 2 are bound}\right) = \left(\frac{x}{1+x}\right)\left(\frac{x^2}{1+x^2}\right).
\]

Now, since binding to the third site can only happen when the first two are bound, we can iterate this argument, leading to (i.e.~by induction),
\[
\label{marginals}
\pi\left(\text{site $i$ is bound}\right) = \frac{x^{2^{i-1}}}{1+x^{2^{i-1}}},
\]
for all $i = 1, \dots, n$, from which we can find the steady-state probability of any occupation state of the sites, including the fully bound state, which has probability
\[
\label{plainNH}
\pi_\mathrm{all} = \pi\left(\text{all sites are bound}\right) =  \prod_{i=1}^n \frac{x^{2^{i-1}}}{1+x^{2^{i-1}}} = \frac{x^{2^n-1}}{\sum_{j=0}^{2^n-1} x^j}.
\]
This expression (and indeed the whole steady-state distribution implied by \eqref{marginals}) is exactly what we would get from \textit{ordered} binding to $2^n-1$ sites---an effect simulated, by this nonequilibrium mechanism, using only $n$ sites!. 

The expression that we get from nested hysteresis, \eqref{plainNH}, saturates \eqref{hypercubebound} only when $\pi_\mathrm{all}$ is very small, and it is not a Hill function. However, we can get one if we now stabilize the two extreme occupation states---the fully bound and totally empty states---by slowing the rates of all transitions leaving them. We can accomplish this by scaling the exit rates from these states by a factor of $q$,  leading to
\[
\label{hillNH}
\pi_\mathrm{all} = \frac{x^{2^n-1}}{1 + q \left(\sum_{j=1}^{2^n-2} x^j\right) + x^{2^n-1}}
\]
which approaches a Hill function with $H = 2^n - 1$ as $q \to 0$, and saturates support bound \eqref{hypercubebound} for all values of $\pi_\mathrm{all}$ simultaneously. A more careful discussion of the construction of kinetic schemes approaching this limit can be found in Appendix \ref{nested_hysteresis_details}.

The possibility of exponential-in-$n$ sensitivity---the optimal nonequilibrium sensitivity achievable with unordered binding to $n$ sites---was missed in the prior numerical work \cite{estrada_information_2016, tran2018precision} which instead seemed to suggest that, e.g.~for $n=3$, $H_\mathrm{eff} \leq 5$. An explicit parametric choice, for the case $n=3$, approaching very close to the Hill function with $H = 2^3-1 = 7$ is illustrated in Figure \ref{fig:nh}(c).

In the same way that sensing and proofreading were two sides of the same coin in the previous section, the results in this section could also be relevant to the question of how highly \textit{specific} gene regulation can be driven by eukaryotic TFs which are characteristically not very specific \cite{wunderlich2009different}. Specificity can be thought of as the sensitivity to the \textit{unbinding} rates \cite{zoller2020nonequilibrium}, which is just the reverse of the sensitivity to the TF concentration, which involves the binding rates. The size of the support in this reversed picture is still $2^n - 1$.

Finally, we note that for any system at thermodynamic equilibrium, the maximum sensitivity of an observable to changing a chemical potential grows linearly with the number of particles, $N$. This fact is closely related to the common intuition that phase transitions can only occur in large systems. The example of nested hysteresis, with sensitivity exponential in $N$, shows that this bound on sensitivity linear in $N$ \textit{dramatically} fails out-of-equilibrium.

\section*{Discussion}

The idea that structure determines function suffuses biology. In the molecular realm, if the conditions of thermodynamic equilibrium prevail, an important aspect of function, sensitivity, is tightly constrained by the most basic structural property---system size. This general physical fact is most familiar in biophysics as the statement that the Hill coefficient for equilibrium binding of a ligand cannot exceed the number of binding sites.

But living things are not at thermodynamic equilibrium, and today at the frontier of molecular biology we are increasingly led to consider this in our models \cite{wong2020gene, zoller2020nonequilibrium}. In this work, we have shown that a structural feature of our models---the size of the support of a perturbation---always limits the sensitivity, at or away from equilibrium. Considering several contexts in which sensitivity is important, we show that our findings unify and extend our understanding of diverse biophysical examples. Table \ref{tablesummary} summarizes these results.

We note that in the nonequilibrium MWC and unordered binding models the size of the support is, up to a fixed additive constant, simply the number of system states---the size of the kinetic scheme. Importantly, it is true for any scheme that the size of the support of any perturbation is less than the number of system states. This means that the effective Hill coefficient is always less than the number of system states. However, in some models, like the KP-like models with a ``fence'', the size of the support may be much less than the number of system states. In all cases we have considered above, the bound in terms of the support is saturatable or ``tight''---it can be approached as closely as desired.

We want to draw attention to three limitations of our work. First, the size of the support is a property of a model and the way it is perturbed, and it may not always correspond to well-defined or observable structure in reality. In our chemotaxis and unordered binding examples, the size of the support is a simple function of the number of binding sites for a ligand, which is usually a relatively concrete notion. However, in the kinetic proofreading and sensing examples, the support consists of some number of states in a model which may be well-distinguished, e.g.~if they correspond to different phosphorylation states of an enzyme, but could also perhaps involve a hard-to-observe conformational ensemble or even continuous degrees of freedom \cite{galstyan2020proofreading}.

Second, some sensitivity mechanisms, like molecular titration \cite{buchler2008molecular} and the Goldbeter-Koshland loop (e.g.~a ``futile'' cycle of phosphorylation and dephosphorylation) \cite{goldbeter1981amplified}, are not well-modeled as small kinetic schemes. This is because they involve the deterministic limit of a reaction network with bimolecular reactions. The support bound still applies (to their master equation), but the presence of enormous symmetries in the transition rates may make the support bound very loose. Also, perturbations involving changing a conserved quantity (e.g.~a total enzyme concentration) in these systems do not fall into the class of perturbations we discuss in this work.

Thirdly, although in several cases we consider, the limit set by the support bound can be saturated---i.e.~we find a mechanism that can achieve it---we do not know how or even when the bound can be saturated in general. One basic situation where the support bound certainly cannot be saturated is when the transitions whose rates depend on $x$ are the forward and reverse transitions between the same two states. Perturbations of such parameters we call ``symmetric edge perturbations'' and we show in \cite{owen2020universal} that the response to such perturbations is bounded by 1, not 2 as the support bound would predict. 

Knowledge of mechanisms and motifs---such as nested hysteresis---that achieve optimal performance subject to natural constraints (e.g.~the number of binding sites) can be extremely useful. This point has been well-argued for by Zoller and others \cite{zoller2021eukaryotic, zoller2020nonequilibrium}, who advocate for a so-called ``normative'' approach, focusing on optimal performance and mechanisms, to tame parametric complexity arising out-of-equilibrium. For any model or model class, the support bound provides a very simple ``first guess'' of what the optimal sensitivity might be---which may be more reliable than the results of numerical search of the parameter space.

\begin{table*}
\begin{center} 
	\begin{tabular}{@{}lccc@{}}
		\toprule
		Model Class     &  Eq. Bound            & Support Bound      & Saturatable? \\
		KP-like, $n$ boundary states   & 1           & $n$   & \checkmark \\ 
		MWC-like, $n$ sites       &  $n$  &  $2n$   &  \checkmark  \\
		Unordered binding, $n$ sites   & $n$          & $2^n - 1$   & \checkmark  \\
	\end{tabular}
\end{center}
\caption{\normalfont Support bound compared to the bounds on sensitivity that hold at thermodynamic equilibrium, for three classes of models we have discussed.}
\label{tablesummary}
\end{table*}

\appendix

                                                                                                                                                                          \section*{Materials and Methods}                                                 

Here we give a proof of the support bound, \eqref{supportbound}. The technical tool we rely on is the Markov chain tree theorem (MTT, also called ``matrix-tree theorem'', see \cite{tutte1948, hill1966, shubert1975, schnakenberg1976network, leighton1986, mirzaev2013laplacian} for details), which gives an explicit algebraic expression for the steady-state distribution $\pi$ of a kinetic scheme in terms of the spanning trees of the associated graph $G$:
\[
\pi_k = \frac{1}{Z} \sum_{\substack{\text{spanning trees of } G \\\text{oriented to k}}}\ \prod_{\text{tree edges $i \to j$}} W_{ji},
\label{mtt}
\]
where $Z$ is the normalization constant, and a \textit{spanning tree} of $G$ is a connected subgraph of $G$ that includes every vertex but has no cycles. In words, the right hand side gives a recipe to find the steady-state probability of a state $k$. It says to consider each spanning tree of $G$ and orient all its edges (choose their direction) so they point towards $k$, which is called the \textit{root} of the tree. Then, for each such oriented tree, multiply together the transition rates associated to all its directed edges. Then, add up the products so formed. The result is proportional to $\pi_k$, up to overall normalization. 

The sum over spanning trees looks forbidding, but to prove the support bound we rely on only two facts, which follow from \eqref{mtt} simply. The first fact is that every term in that sum is a \textit{positive} monomial, being a product of nonzero transition rates. This means the ratio of any positive observables is a ratio of polynomials in $x$:
\begin{equation}\label{method}
\frac{\langle A \rangle_\pi}{\langle B \rangle_\pi} = \frac{\sum_{i = a_\mathrm{min}}^{a_\mathrm{max}} k_{i-a_\mathrm{min}} x^i }{\sum_{j = b_\mathrm{min}}^{b_\mathrm{max}} q_{j-b_\mathrm{min}} x^j} = x^{a_\mathrm{min} - b_\mathrm{min}} \frac{\sum_{i = 0}^{a_\mathrm{max}-a_\mathrm{min}} k_i x^i }{\sum_{j = 0}^{b_\mathrm{max}-b_\mathrm{min}} q_j x^j} 
\end{equation}
where $k_i$ and $q_j$ are positive quantities that do not depend on $x$, and $a_\mathrm{max}$, $a_\mathrm{min}$, $b_\mathrm{max}$, and $b_\mathrm{min}$ are nonnegative integers.

Differentiating this expression we find
\begin{equation}
\begin{split}
\frac{d \log \langle A \rangle_\pi / \langle B \rangle_\pi}{d \log x} &= \\
\left(a_\mathrm{min} - b_\mathrm{min}\right) + &\left(\frac{\sum_{i = 0}^{a_\mathrm{max}-a_\mathrm{min}} i k_i x^i }{\sum_{i = 0}^{a_\mathrm{max}-a_\mathrm{min}}  k_i x^i } - \frac{\sum_{j = 0}^{b_\mathrm{max}-b_\mathrm{min}} j q_j x^j }{\sum_{j = 0}^{b_\mathrm{max}-b_\mathrm{min}} q_i x^j }\right).
\end{split}
\end{equation}
The second term in brackets is not less than $-b_\mathrm{max} + b_\mathrm{min}$ and not more than  $a_\mathrm{max} - a_\mathrm{min}$, so we get
\begin{equation}
a_\mathrm{min} - b_\mathrm{max} \leq  \frac{d \log \langle A \rangle_\pi / \langle B \rangle_\pi}{d\log x} \leq a_\mathrm{max} - b_\mathrm{min}.
\end{equation}

The second fact about \eqref{mtt} that we need is that each oriented spanning tree has at most one directed edge emanating from each vertex (it has none coming of its root). Recalling the definition of support, it follows that each monomial in \eqref{mtt} picks up at most $m$ factors of $x$, where $m$ is the size of the support of the perturbation. This means $a_\mathrm{max}$ and $b_\mathrm{max}$ are both no greater than $m$, which leads to
\begin{equation}\label{supportbound2}
\left|\frac{d \log \langle A \rangle_\pi / \langle B \rangle_\pi}{d \log x}\right| \leq m.
\end{equation}

\quad

\section*{Acknowledgments}

J.A.O thanks Leonid Mirny for advice and support.

\bibliographystyle{unsrtnat}
\bibliography{support_refs_v2}

\clearpage

\appendix
\onecolumngrid

\section{Proof that $H_\mathrm{eff} = \log \left(81\right) / \log(S_{0.9}/S_{0.1})$ is bounded by size of support}
\label{global_hill_bounded}
In this section, we will see how the size of the support bounds the effective Hill coefficient, when it is defined according to \eqref{hillcoeff81} (in the main text, reproduced here),
\[
H_\mathrm{eff} = \frac{\log \left(81\right)}{\log(S_{0.9}/S_{0.1})}.
\]

Suppose we are interested in the sensitivity properties of a function $f(x)$ which is positive, monotonically increasing in the parameter $x$, and bounded above by a value $f_\mathrm{max}$. These assumptions are effectively required to be able to apply (\ref{hillcoeff81}), e.g.~because the definition presupposes the existence and uniqueness of the values $S_{0.1}$ and $S_{0.9}$, which, recall, are the values of the input variable $x$ for which $f(x)$ achieves $10\%$ and $90\%$, respectively, of its maximum range. We additionally suppose that $f(x)$ is a positive observable of some kinetic scheme, which implies that $f_\mathrm{max} - f(x)$ is a positive observable as well. 

The support bound then gives
\[
\frac{d \log \left(\frac{f(x)}{f_\mathrm{max}-f(x)}\right) }{d \log x} \leq m,
\]
where $m$ is the size of the support of the perturbation of $x$. Note that since $f(x)$ is increasing the left hand side is positive. Then, define $z = \log x$ (we assume, as we do throughout this work, that $x > 0$), and write $z_{0.1} = \log S_{0.1}$ and $z_{0.9} = \log S_{0.9}$. Now we integrate the inequality
\[
\int_{z_{0.1}}^{z_{0.9}} \frac{d \log \left(\frac{f(x)}{f_\mathrm{max}-f(x)}\right) }{d z} \, d z \leq \int_{z_{0.1}}^{z_{0.9}} m \, d z,
\]
which yields
\[
\log \left(\frac{0.9 f_\mathrm{max}}{f_\mathrm{max}- 0.9 f_\mathrm{max}}\right) - \log \left(\frac{0.1 f_\mathrm{max}}{f_\mathrm{max}-0.1 f_\mathrm{max}}\right)  \leq m \left(z_{0.9} - z_{0.1}\right)
\]
\[
\log \left(0.9/0.1\right) - \log \left(0.1/0.9\right) \leq m \left(z_{0.9} - z_{0.1}\right)
\]
from which it follows that $\log\left( 81\right) \leq m \log(S_{0.9}/S_{0.1})$, which implies $H_\mathrm{eff} \leq m$, for the definition (\ref{hillcoeff81}) of $H_\mathrm{eff}$, as desired.

\section{Proof of equilibrium equality (\ref{eqbound}) directly from detailed balance}
\label{proof_eqbound}
Here, our goal is prove that for any detailed balanced kinetic scheme, the equality
\[
\frac{d \log \pi_X}{d \log x} =\left[ \langle n_b \rangle_X - \langle n_b \rangle_{\overline{X}} \right] (1-\pi_X), 
\]
holds. This is Eq.~\eqref{eqbound} from the main text. Recall the $X$ is a set of system states, $\overline{X}$ is the set of system states not in $X$, and $\pi_X = \sum_{i \in X} \pi_i$.

We begin by expressing the steady-state probability of a state $i$ in terms of ratios of steady-state probabilities, using simply the normalization of probability, $\sum_j \pi_j = 1$, to get 
\[
\pi_i = \frac{1}{1+\sum_{j\neq i} \frac{\pi_j}{\pi_i}}.
\]
For any detailed balance scheme, the ratio of the probabilities of two states $i$ and $j$ depends only on the ratio of transition rates along a path $i \to 1 \to 2 \to \cdots k \to j$ connecting them, as follows:
\[
\label{pathratio}
\frac{\pi_j}{\pi_i} = \frac{W_{1i}}{W_{i1}} \frac{W_{21}}{W_{12}} \cdots \frac{W_{jk}}{W_{kj}}.
\]
If there are multiple paths between $i$ and $j$, this product of ratios will be the same for each path (this is exactly equivalent to the condition of detailed balance), so we may restrict attention to a single one. Now suppose $i$ represents a system state with $n_i$ ligands bound, and $j$ represents a system state with $n_j$ ligands bound. If along the (directed) path from $i$ to $j$, $p$ transitions represent binding of the ligand and along the (directed) reverse path from $j$ to $i$, $q$ transitions represent binding, then $p - q = n_j - n_i$. 

Now we make our key assumption, which is the mass-action assumption that the transitions representing binding have rates linear in the ligand concentration $x$. We also assume that all other transition rates are independent of $x$. This means we can write
\[
\pi_i = \frac{1}{1 + \sum_{j \neq i} x^{n_j-n_i} A_{ji}},
\]
where $A_{ji}$ is a positive number depending on transition rates but independent of $x$.

We are interested in fact in the steady-state probability of some set of states $X$, which we can now write as:
\[
\pi_X = \sum_{i \in X} \pi_i = \sum_{i \in X}  \frac{1}{1 + \sum_{j \neq i} x^{n_j-n_i} A_{ji}}.
\]

The logarithmic sensitivity of $\pi_X$ to changes in $x$ can now be computed by differentiating:
\[
\frac{\partial \log \pi_X}{\partial \log x} = -\frac{1}{\pi_X} \sum_{i \in X}\left( \pi_i^2 \sum_{j \neq i} \left(n_j-n_i\right)x^{n_j-n_i} A_{ji}\right) = -\frac{1}{\pi_X} \sum_{i \in X}\left( \pi_i^2 \sum_{j} \left(n_j-n_i\right) \frac{\pi_j}{\pi_i}\right),
\]

\[
\frac{\partial \log \pi_X}{\partial \log x} = -\frac{1}{\pi_X} \sum_{i \in X} \pi_i \left( \sum_{j} \left(n_j-n_i\right) \pi_j\right) = -\frac{1}{\pi_X} \sum_{i \in X}\pi_i \left(\sum_{j\in X} \left(n_j-n_i\right) \pi_j + \sum_{j\notin X} \left(n_j-n_i\right) \pi_j\right),
\]
now we do the sums over $j$:
\[
\frac{\partial \log \pi_X}{\partial \log x} = -\frac{1}{\pi_X} \sum_{i \in X} \pi_i \left( \pi_X \langle n_b\rangle_X - n_i \pi_X + \pi_{\overline{X}} \langle n_b\rangle_{\overline{X}} - n_i \pi_{\overline{X}} \right),
\]
and those over $i$, yielding the result, Eq.~(\ref{eqbound}):
\[
\frac{\partial \log \pi_X}{\partial \log x} = -\left(\pi_{\overline{X}} \langle n_b \rangle_{\overline{X}} - \pi_{\overline{X}} \langle n_b \rangle_X \right) = \left(\langle n_b \rangle_{X} - \langle n_b \rangle_{\overline{X}} \right) (1-\pi_X).
\]

What goes wrong in this argument when detailed balance is broken? In that case, the ratio $\pi_j/\pi_i$ depends on the transition rates not according to Eq.~\eqref{pathratio}, but in a more complicated way given in general by the Markov chain tree theorem, as described in the main text. The ratio $\pi_j/\pi_i$ is no longer a homogeneous function of $x$, but is a quotient of polynomials in $x$ (see Eq.~(\ref{method})).   

\section{Application of the equilibrium equality (\ref{eqbound}) to the findings of Fukuoka et al.}
\label{eqbound_fukuoka}
In the experiment of Fukuoka et al. \cite{fukuoka_direct_2014}, \textit{E. coli} cells expressing CheY-GFP fusion protein were tethered to a substrate by one of their flagellar filaments, enabling simultaneous observation of rotational direction (the whole cell body rotates) and CheY binding to the motor, which can be quantified by the intensity of a fluorescent CheY-GFP spot at the center of rotation. The authors observe spontaneous directional switching and report the average numbers of CheY-P bound to the motor when it rotates clockwise and when it rotates counterclockwise.

The spontaneous directional switching of different motors on the same cell has also been observed to be synchronized \cite{terasawa2011coordinated}, which suggests that the switching may be due to spontaneous (intrinsic) fluctuations in the CheY-P concentration in the cell. It might be, then, that the numbers of bound CheY-P reported by Fukuoka et al.~should be thought of as measurements at two slightly different CheY-P concentrations $x_-$ and $x+$, with $x_- < x_+$. By contrast, the right hand side of Eq.~(\ref{eqbound}) involves the difference of the number bound
$\langle n_b \rangle_\mathrm{CW}(x) - \langle n_b \rangle_\mathrm{CCW}(x)$ as a \textit{fixed} concentration $x$.

However, for any detailed balanced model, the mean number bound in each state $\langle n_b \rangle_\mathrm{CW}(x)$ and $\langle n_b \rangle_\mathrm{CCW}(x)$ are increasing functions of $x$. This means that for any $x_- < x < x_+$,
\[
\langle n_b \rangle_\mathrm{CW}(x) - \langle n_b \rangle_\mathrm{CCW}(x) < \langle n_b \rangle_\mathrm{CW}(x_+) - \langle n_b \rangle_\mathrm{CCW}(x_-), 
\]
and that the measurement $11 = 13 - 2$ of Fukuoka et al.---the right hand side of this inequality---should exceed the effective Hill coefficient, in any detailed balanced scheme.

\section{Saturation of the nonequilibrium MWC bound, (\ref{chemotactic_bound})}
\label{chemotaxis_sat}
Here we show that the bound \eqref{chemotactic_bound} can be approached arbitrarily closely in an appropriate limit of transition rates, and that in fact, $\pi_\mathrm{CW}$ can be made to approach a Hill function with $H = 2n$.

Suppose that binding and unbinding of CheY-P is very fast compared to the $\mathrm{CCW} \leftrightarrow \mathrm{CW}$ switching transitions. Suppose also that CheY-P binds at equilibrium with dissociation constant depending on the rotation state, $K_\mathrm{CW}$ or $K_\mathrm{CCW}$, independently (non-cooperatively) to each binding site. Finally, suppose that the $\mathrm{CCW} \to \mathrm{CW}$ transition can only occur when $n$ CheY-P molecules are bound, and the $\mathrm{CW} \to \mathrm{CCW}$ transition can only occur when none are bound. Suppose that when they can occur, these directional switching transitions occur at rate $r$. 
 
Under these assumptions, the penultimate of which breaks detailed balance, the effective rate of the $\mathrm{CCW} \to \mathrm{CW}$ transition is given by $r$ times the probability of $n$ molecules being bound, given that the rotation state is $\mathrm{CCW}$, which is $x^n/\left(K_\mathrm{CCW}+x\right)^n$. Similarly, the effective rate of the $\mathrm{CW} \to \mathrm{CCW}$ transition is given by $r$ times the probability of no molecules being bound, given that the rotation state is $\mathrm{CW}$, which is $K_\mathrm{CW}^n/\left(K_\mathrm{CW}+x\right)^n$. Therefore, the steady-state clockwise bias is given by
\[
\label{mwc_optimizer}
\pi_\mathrm{CW}(x) \approx \frac{\left(\frac{x}{K_\mathrm{CCW}+x}\right)^n}{\left(\frac{x}{K_\mathrm{CCW}+x}\right)^n+\left(\frac{K_\mathrm{CW}}{K_\mathrm{CW}+x}\right)^n} = \frac{1}{1+\left(\frac{K_\mathrm{CW}}{K_\mathrm{CW}+x}\right)^n \left(\frac{K_\mathrm{CCW}+x}{x}\right)^{n}},
\] 
where the approximate equality relies on the timescale separation mentioned above. Careful discussion and justification of this kind of approximation can be found in, e.g.~Simon and Ando (1961) \cite{simon1961aggregation} and Courtois (1975) \cite{courtois1975error}. In that literature, a system satisfying the kind of timescale separation assumptions described above is called \textit{nearly completely decomposable}. See also Haken (1983), pp.~204--205 \cite{haken1983introduction}.  

Now we take $K_\mathrm{CCW} = K/\epsilon$ and $K_\mathrm{CW} = K\epsilon$, so that
\[
\pi_\mathrm{CW}(x) = \frac{1}{1+\left(\frac{K\epsilon}{K\epsilon+x}\right)^n \left(\frac{K/\epsilon+x}{x}\right)^{n}} = \frac{1}{1+\left(\frac{K}{x}\right)^n \left(\frac{K+x\epsilon}{x+K\epsilon}\right)^{n}}.
\] 
And so if we send $\epsilon \to 0$, so that the clockwise state has a much higher affinity for CheY-P $K_\mathrm{CW} \ll x \ll K_\mathrm{CCW}$, we find that $\pi_\mathrm{CW}$ approaches a Hill function with $H = 2n$:
\[
\pi_\mathrm{CW}(x) \to \frac{x^{2n}}{K^{2n}+x^{2n}},
\] 
which saturates \eqref{chemotactic_bound}.

If instead $K_\mathrm{CCW} = K_\mathrm{CW} = K$, then \eqref{mwc_optimizer} becomes
\[
\pi_\mathrm{CW}(x) \approx \frac{x^{n}}{K^{n}+x^{n}},
\] 
while the average number of ligands bound is the same in the clockwise and counterclockwise states. 

\section{Exponential sensitivity by nested hysteresis}
\label{nested_hysteresis_details}
In this section, we show that the mechanism of \textit{nested hysteresis}, described in the main text, can achieve a sensitivity of $2^n - 1$, saturating the support bound for unordered binding. We also show that a simple modification of this mechanism---stabilizing the fully bound and fully unbound states by scaling the rates of their exit transitions---results in the probability of the fully bound state $\pi_\mathrm{all}(x)$, viewed as a function of the ligand concentration $x$, approaching as closely as desired a Hill function with $H = 2^n - 1$:
\[
\pi_\mathrm{all}(x) \to \frac{x^{2^n-1}}{1+x^{2^n-1}}.
\]
To lighten notation, we will from now on suppress functional dependence on $x$, e.g. writing $\pi_\mathrm{all}$ instead of $\pi_\mathrm{all}(x)$. We will also write $\pi_\mathrm{none}$ for the steady-state probability that no ligands are bound.

Our arguments will only establish convergence \emph{pointwise} in $x$, which especially for this second claim---about the approach to a Hill function---may feel unsatisfactory, since a sequence of functions converging pointwise may never look similar to their limit function. Therefore, we supplement these arguments with a \textit{Mathematica} notebook ({\tt https://github.com/jaowen/nested-hysteresis}) with code to explicitly construct, for any $n$, kinetic schemes as close as desired to the limiting case we discuss here. In the special case of $n = 3$, we also provide a computer-assisted verification of an inequality that implies uniform convergence to a Hill function with $H = 7$.

The first step in our argument is to describe a nonequilibrium mechanism to double the sensitivity of any kinetic scheme, effectively by coupling it to a new binary degree of freedom in a special way---doubling its state space. Then, by iterating this construction, we will be able to build up to nested hysteresis.

\subsection{Sensitivity doubling}

Suppose we have a kinetic scheme with graph $G$, whose transition rates may depend (linearly) on a parameter $x$, and in which we distinguish two states, which we call $a$ and $b$. Define 
\[
\tilde{\pi}_a =\sum_{\substack{\text{spanning trees of } G \\\text{oriented to a}}}\ \prod_{\text{tree edges $i \to j$}} W_{ji}
\]
to be the sum of oriented spanning tree weights given by the Markov chain tree theorem (MTT, see discussion in ``Materials and Methods'' of main text) for the probability of state $a$, \emph{prior to normalization}. $\tilde{\pi}_b$ is defined analogously, but with trees rooted at state $b$. We also need to define another quantity, which involves \emph{spanning 2-forests} of $G$. A spanning 2-forest of $G$ is a subgraph of $G$ with two connected components, containing every vertex, and having no cycles. The components of spanning 2-forests are trees (graphs with no cycles). If $a$ and $b$ are contained in different components of the spanning forest, then we can orient the forest so that one component tree is rooted at $a$ and the other at $b$. We define $F_{ab}$ to be the sum of the weights of all such forests, oriented in that way:

\[
F_{ab} =\sum_{\substack{\text{spanning 2-forests of } G \\\text{oriented to a, b}}}\ \prod_{\text{tree edges $i \to j$}} W_{ji}
\]

Now suppose we construct a new scheme whose graph $G'$ is formed by joining together two copies of $G$ as follows:
\begin{center}
	\scalebox{0.4}{\includegraphics{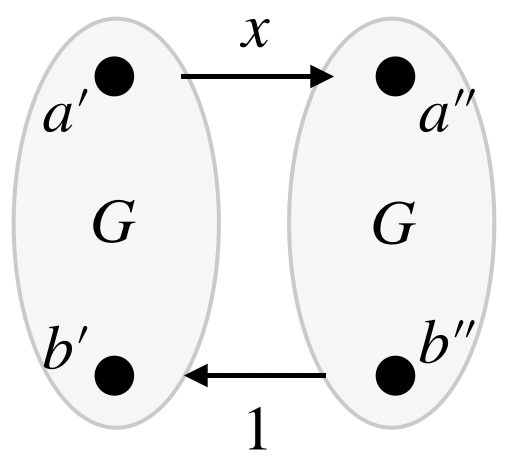}}
\end{center}
where in the copies of $G$ we have renamed the states $a$ and $b$, to distinguish each from its counterpart in the other copy of $G$ and in the original scheme. 

Our goal now will be to find, for this new kinetic scheme, the steady-state ratio $\pi_{a''}/\pi_{b'}$. We will do this by applying the MTT, starting with spanning trees rooted at $a''$. There are two kind of such trees. The first kind consists of two spanning trees of $G$ (one for each copy) rooted at $a'$ and $a''$, plus the edge with rate $x$. The second kind consists of a spanning tree of $G$ rooted at $a'$, a spanning 2-forest with components rooted at $a''$ and $b''$, and the edges with rates $x$ and $1$. These two classes of spanning trees of $G'$, plus their contributions (in terms of $\tilde{\pi}_a$, $F_{ab}$ and $x$) to the sum in the MTT, are illustrated below:
\begin{center}
	\scalebox{0.4}{\includegraphics{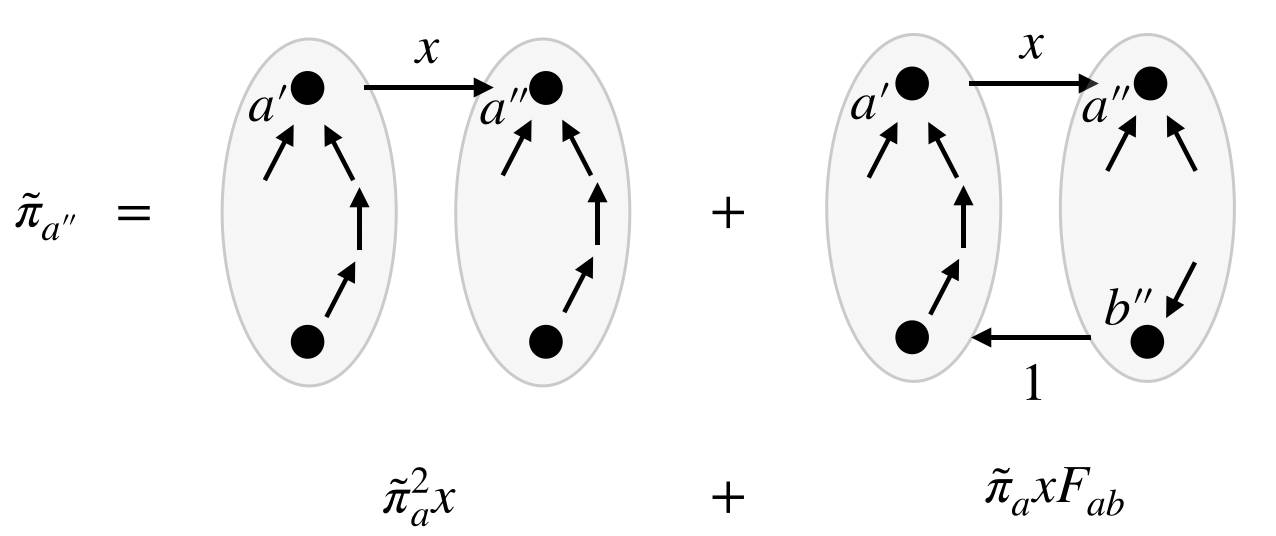}}
\end{center}

Reasoning similarly about the spanning trees of $G'$ rooted at $b'$, we get:
\begin{center}
	\scalebox{0.4}{\includegraphics{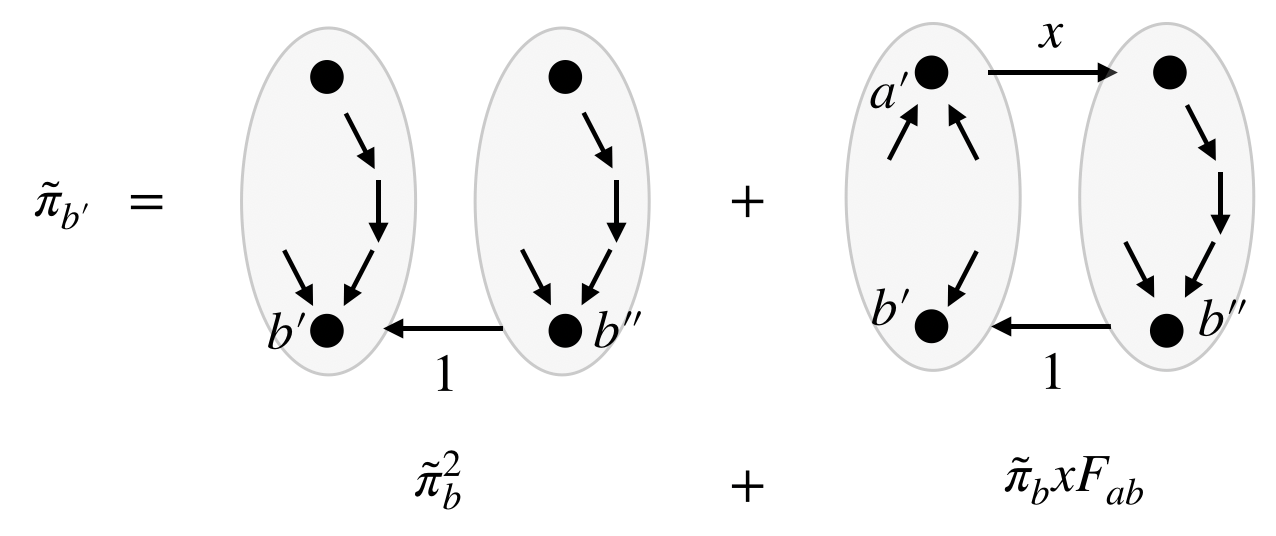}}
\end{center}

Putting these together, we get from the MTT that the steady-state probability ratio is given by
\[
\label{messyratio}
\frac{\pi_{a''}}{\pi_{b'}} = \frac{\tilde{\pi}_{a''}}{\tilde{\pi}_{b'}} = \frac{\tilde{\pi}_a^2 x + \tilde{\pi}_a x F_{ab}}{\tilde{\pi}_b^2 + \tilde{\pi}_b x F_{ab}}.
\]

Now we are ready to introduce the timescale separation that is critical to this mechanism. We do this by supposing that all the transition rates inside the copies of $G$ (e.g.~inside the gray ovals in the graphs of $G'$) are scaled by the same factor $s$, which we will then send towards infinity. The key fact is that the terms in \eqref{messyratio} are homogeneous in this scale factor $s$. This is because a spanning tree of a graph with $N$ vertices always has $N-1$ edges, and a spanning 2-forest has $N-2$. So if $G$ has $N$ vertices, then we get
\[
\label{rat}
\frac{\pi_{a''}}{\pi_{b'}} = \frac{(s^{N-1})^2\tilde{\pi}_a^2 x + (s^{N-1})(s^{N-2}) \tilde{\pi}_a x F_{ab}}{(s^{N-1})^2\tilde{\pi}_b^2 + (s^{N-1})(s^{N-2})\tilde{\pi}_b x F_{ab}}=\frac{s\tilde{\pi}_a^2 x +  \tilde{\pi}_a x F_{ab}}{s\tilde{\pi}_b^2 + \tilde{\pi}_b x F_{ab}}.
\]

In the limit $s \to \infty$ we then get 
\[
\frac{\pi_{a''}}{\pi_{b'}} \to x \left(\frac{\tilde{\pi}_a}{\tilde{\pi}_b}\right)^2 = \left(\frac{\pi_a}{\pi_b}\right)^2,
\]
pointwise in $x$. Notably, it is also true that the derivative
\[
\frac{\partial}{\partial x} \left(\frac{\pi_{a''}}{\pi_{b'}}\right) \to \frac{\partial}{\partial x} \left( x \left(\frac{\pi_a}{\pi_b}\right)^2\right),
\]
pointwise, as $s \to \infty$. This is false in general (pointwise convergence of a sequence of functions does not imply pointwise convergence of derivatives to the derivative of the limit) but is true for this rational function \eqref{rat} as a consequence of it being true for polynomials. An argument that it is true for polynomials can be found at \cite{stackexchange}. Now to see that it is true for \eqref{rat}, write $p(x,s) = \tilde{\pi}_a^2 x +  \tilde{\pi}_a x F_{ab}/s$ and $q(x, s) = \tilde{\pi}_b^2 + \tilde{\pi}_b x F_{ab}/s$. These are both polynomials in $x$ and they converge pointwise to limits $p(x) = \tilde{\pi}_a^2 x$ and $q(x) = \tilde{\pi}_b^2$, respectively, as $s \to \infty$. Now consider $\tilde{\pi}_{a''}/\tilde{\pi}_{b'} = p(x, s) / q(x, s)$. The derivative with respect to $x$ is
\[
\label{deriv}
\frac{\partial}{\partial x} \left(\frac{\tilde{\pi}_{a''}}{\tilde{\pi}_{b'}}\right) = \frac{\left(\partial_x p(x,s)\right) q(x,s) - p(x,s) \left(\partial_x q(x,s)\right)}{q(x,s)^2}.
\]
Note that as $s \to \infty$, the derivatives $\partial_x p(x,s) \to \partial_x p(x)$ and $\partial_x q(x,s) \to \partial_x q(x)$ (because $p$ and $q$ are polynomials in $x$, \cite{stackexchange}), and so the right hand side of \eqref{deriv} converges pointwise to the derivative with respect to $x$ of $p(x)/q(x) = \tilde{\pi}_a^2 x / \tilde{\pi}_b^2$, as desired.

These arguments apply equally well to the \textit{logarithmic} derivative, and so the construction described in this section can be thought of as doubling (and then adding one to) a logarithmic sensitivity. To see this, suppose for the original graph $G$ we have $\pi_a / \pi_b = x^m$ for some $m$, so that $\partial \log(\pi_{a} / \pi_{b}) / \partial \log x = m$. Then the arguments above show that, in the new scheme, $\partial \log(\pi_{a''} / \pi_{b'})/\partial \log x \to 2 m + 1$ as $s \to \infty$. Note that the recurrence $m_1 = 1$, $m_{i+1} = 2 m_i + 1$ has solution $m_n = 2^n - 1$.

\subsection{Iterative construction}

We can repeat the whole construction described in the previous section, starting with the graph $G'$ instead of $G$, and taking $a = a''$, $b = b'$. The iteration of this procedure---starting from a kinetic scheme with two states representing binding and unbinding of a ligand to a single site---is what is depicted in Figure \ref{fig:unordered}(c) of the main text:
\begin{center}
	\scalebox{0.5}{\includegraphics{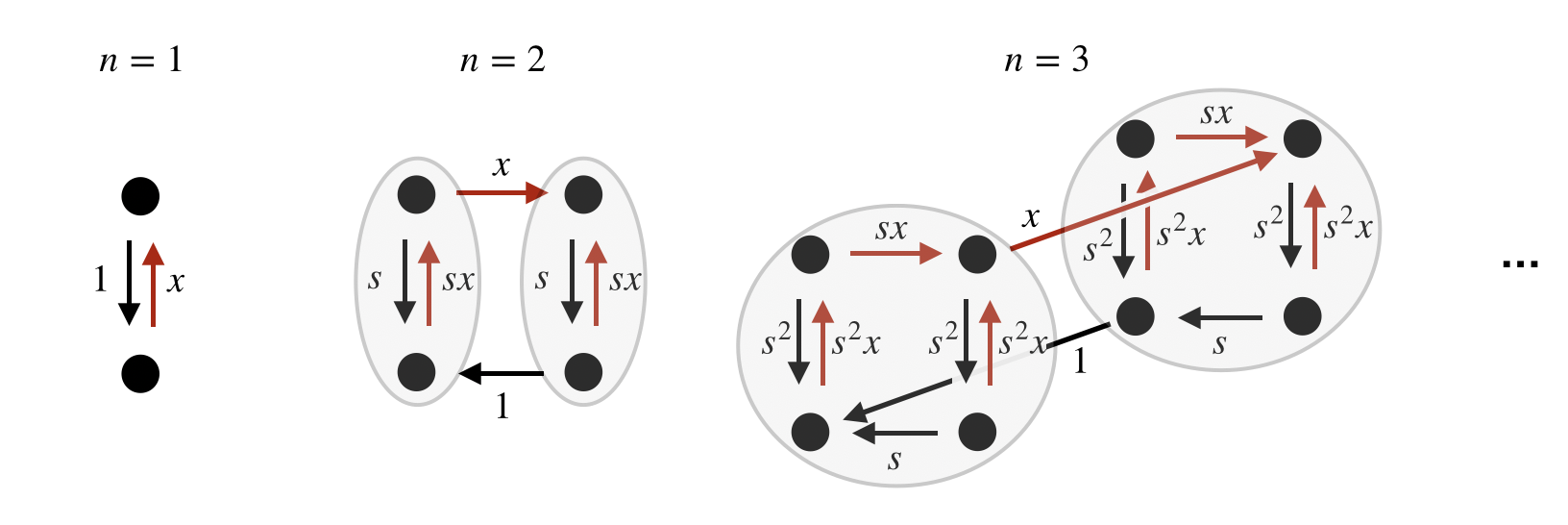}}
\end{center}

To be explicit, let's denote each possible binding state by a string of zeros and ones indicating which sites are occupied. For example, when there is a single binding site $n=1$, the two states are `0' (unoccupied) and `1' (occupied). When there are two binding sites, the states are `00', `10', `01', and `11', and so on for higher numbers of binding sites. In the case $n=1$ the states $a$ and $b$ (for the purposes of the sensitivity doubling described in the last section) are $a = 0$ and $b = 1$. When constructing $n=2$, we have $a'' = 11$ and $b' = 00$ and these become the new $a$ and $b$ for the next step, and so on.

In the base case ($n = 1$), the ratio of probabilities $\pi_a / \pi_b = \pi_1/\pi_0$ equals $x$. Then, one step of sensitivity doubling constructs a scheme (the case $n = 2$) where the ratio of two state probabilities $\pi_{11} / \pi_{00} \to x^3$ as $s \to \infty$ (pointwise). Let's define $f(x, s) \equiv \pi_{11} / \pi_{00}$. 

For any finite value of $s$, $f(x,s)$ won't quite be $x^3$---does this undermine our ability to ``nest'' the argument and build up to a scheme which approaches $x^{2^n-1}$ as $s \to \infty$? It turns out that it does not. To see this, let's be very explicit, considering the next step of the iteration. We must allow that the new scale factor introduced might not take the same value as the one (called $s$) in the first step, so we will call the new one $s'$. Write $g(x, s, s') \equiv  \pi_{111} / \pi_{000}$. Now, our sensitivity doubling arguments give $\lim_{s\to\infty} f(x,s) = x^3$ and $\lim_{s'\to\infty} g(x,s,s') = x f(x,s)^2$. It turns out this does imply that $\lim_{s \to \infty} g(x,s,s) = x(x^3)^2 = x^7$. To see this, note that for any $\epsilon$ there is a number $S_1$ such that $s > S_1$ guarantees $|x f(x,s)^2 - x^7| \leq \epsilon/2$ (by the first limit and continuity of $x f^2$ in $f$), and another number $S_2$ such that $s' > S_2$ guarantees $|g(x,s,s') - x f(x,s)^2| \leq \epsilon/2$. And so for any $\epsilon$, choosing $s > \max(S_1, S_2)$ gives $|g(x,s,s) - x^7| \leq \epsilon$. Note that $S_1$ and $S_2$ may depend on $x$.

To summarize, the arguments we have given so far allow us to construct, for any $n$, a scheme of ligand binding with rates depending on a parameter $s$ such that 
\[
\frac{\pi_\mathrm{all}}{\pi_\mathrm{none}} \to x^{2^n-1}
\]
as $s \to \infty$.

\subsection{Stabilizing extreme states to get a Hill function}

Finally, we will show here how, given a kinetic scheme with any given value for the steady-state ratio of two state probabilities $\pi_i/\pi_j = \alpha > 0$, we can create a scheme with $\pi_i$ as close as desired to $\alpha/\left(1+\alpha\right)$. Importantly, the construction does not depend on the value of $\alpha$.

We begin by defining, for any $k$,
\[
\tilde{\pi_k} =\sum_{\substack{\text{spanning trees of } G \\\text{oriented to k}}}\ \prod_{\text{tree edges $i \to j$}} W_{ji}
\]
to be the sum of spanning tree contributions established by the MTT to be proportional to the steady-state probability $\pi_k$. Note that we have $\tilde{\pi_i}/\tilde{\pi_j} = \pi_i/\pi_j = \alpha$, and
\[
\label{ijscale}
\pi_i = \frac{\tilde{\pi_i}}{\sum_k \tilde{\pi_k}} = \frac{\tilde{\pi_i}}{\tilde{\pi_i}+\tilde{\pi_j}+\sum_{k\neq i,j} \tilde{\pi_k}}.
\]

Now suppose we scale all the rates of transitions \textit{leaving} $i$ and $j$ by a factor $q$, pictured schematically below,
\begin{center}
	\scalebox{0.5}{\includegraphics{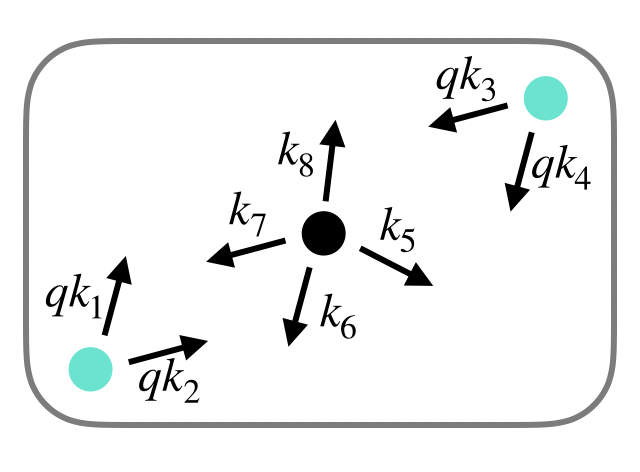}}
\end{center}
where the states $i$ and $j$ are labeled in cyan.

Any spanning tree oriented to a state \textit{other than} $i$ and $j$ must include \textit{exactly one} directed edge leaving $i$ and one leaving $j$,  which means two factors of $q$. Those oriented to $i$ or $j$, by contrast, pick up only one factor of $q$ (for leaving $j$, or $i$, respectively). This means that scaling by $q$ changes \eqref{ijscale} to
\[
\pi_i = \frac{q\tilde{\pi_i}}{q\tilde{\pi_i}+q\tilde{\pi_j}+q^2\sum_{k\neq i,j} \tilde{\pi_k}} = \frac{\tilde{\pi_i}}{\tilde{\pi_i}+\tilde{\pi_j}+q\sum_{k\neq i,j} \tilde{\pi_k}}.
\]
And so, in the limit $q \to 0$, we find
\[
\pi_i \to \frac{\tilde{\pi_i}}{\tilde{\pi_i}+\tilde{\pi_j}} = \frac{\alpha}{1+\alpha},
\] 
as desired.

Applying this ``stabilization of extremes'' to nested hysteresis finally gives
\[
\lim_{q \to 0} \lim_{s \to \infty} \pi_\mathrm{all} = \lim_{q\to 0} \frac{x^{2^n-1}}{1 + x^{2^n-1} + q \left(\sum_{j=1}^{2^n-2} x^j\right)} = \frac{x^{2^n-1}}{1 + x^{2^n-1}}.
\]

\subsection{The cases $n=2$ and $n=3$}

For illustrative purposes, we give here explicit expressions for the probability of the fully bound state $\pi_\mathrm{all}$, in nested hysteresis with stabilized extremes, in the cases $n = 2$ and $n = 3$. We choose $q = 1/s$, but leave $s$ as a parameter. It turns out that this choice of $q$ allows us to describe (for $n=2, 3$) the limiting procedure leading a Hill function with $H = 2^n = 1$ as a single limit $s \to \infty$. Note that we do \textit{not} prove that this choice of $s$ dependence for $q$ works for any $n$. Instead our arguments above only establish that the iterated limit, first of $s \to \infty$ and then of $q \to 0$, leads to a Hill function with $H = 2^n = 1$.

\[
\pi_\mathrm{all} = \pi_{11} = \frac{x^2 (s x+1)}{s \left(x^3+1\right)+2 x (x+1)}
\]

\[
\resizebox{.9\hsize}{!}{$\pi_\mathrm{all} = \pi_{111} = \frac{x^3 (s x+1)^2 \left(s^2 x^2+s+x\right)}{s^4
   \left(x^7+1\right)+s^3 x (x+1) \left(x^2+x+1\right) (x (3
   x-4)+3)+s^2 x (x+1) \left(x^2+x+1\right)^2+3 s x^2 (x+1)
   \left(x^2+1\right)+2 x^3 (x+1)}$}
\]
The code used to generate these expressions can be found in the supplemental \textit{Mathematica} notebook, available online at {\tt https://github.com/jaowen/nested-hysteresis}. The same code can be used to generate explicit expressions at least up to $n = 5$---they become very large but \textit{Mathematica} confirms the correct limit as $s \to \infty$, e.g.~in the case $n=5$ the limit being $x^{31}/(1+x^{31})$.

\end{document}